\documentclass[12pt,prc,preprint,showpacs,showkeys,nobibnotes]{revtex4-1}
\usepackage{color}
\usepackage{amsmath}
\usepackage{graphicx}
\usepackage{amssymb}
\usepackage{esint}

\renewcommand{\vec}[1]{\boldsymbol{#1}}

\DeclareMathOperator{\STr}{Str}
\renewcommand{\Re}{\mathrm{Re}}
\renewcommand{\Im}{\mathrm{Im}}

\begin{document}

\title{Functional renormalisation group for few-nucleon systems:\\
 SU(4) symmetry and its breaking}

\author{Michael C. Birse, Boris Krippa, Niels R. Walet}

\affiliation{Theoretical Physics Division, School of Physics and Astronomy, 
University of Manchester, Manchester,
M13 9PL, UK}

\date{\today}
\begin{abstract}
We apply the functional renormalisation group to few-nucleon systems.
Our starting point is a local effective action that includes three-
and four-nucleon interactions, expressed in terms of nucleon and
two-nucleon boson fields. The evolution of the coupling
constants in this action is described by a renormalisation group flow.
We derive these flow equations both in the limit of exact Wigner SU(4)
symmetry and in the realistic case of broken symmetry.  In the
symmetric limit we find that the renormalisation flow equations decouple, and can be
combined into two sets, one of which matches the known results for bosons, and the other 
result matches the one for 
a single flavour of spin $1/2$ fermions.
The equations show
universal features in the unitary limit, which is obtained when the
two-body scattering length tends to infinity.  We calculate the spin-quartet neutron-deuteron
scattering length and the deuteron-deuteron scattering lengths in the
spin-singlet and quintet channels.
\end{abstract}



\maketitle

\section{Introduction}

\global\long\def\ivc{\vec{\ib}}
\global\long\def\vvc{\vec{\vb}}
\global\long\def\vb{t}
\global\long\def\ib{s}
\global\long\def\isum{\ib}
\global\long\def\vsum{\vb}
\global\long\def\fermion{N}
\global\long\def\boson{D}
\global\long\def\pB{\mathcal{E}_{D}}
\global\long\def\lambdahh{\lambda^{(1/2,1/2)}}
\global\long\def\lambdath{\lambda^{(3/2,1/2)}}
\global\long\def\lambdaht{\lambda^{(1/2,3/2)}}
\global\long\def\ed{\mathcal{E}_{d}}

Few-body physics provides a solid starting point for analysing
various nonperturbative approaches and methods which can then be used
for studying more complicated many-body problems. The study of
few-body systems also allows us to determine the input parameters
needed for those complex problems, without the need to rely on further
approximations. One promising method is the Functional Renormalisation
Group (FRG) \cite{Wet,Morr}, as it provides a framework that can
simultaneously describe both few- and many-body systems. Some reviews of
the method and its applications can be found in
Refs.~\cite{Ber,Del,Paw}.

The approach is based on a running effective action (REA) which is a
generalisation of the standard quantum mechanical effective action --
the generating functional of the one-particle irreducible Green
functions. The REA includes the effects of all the fluctuations with
momenta in the region $q^{2}\gtrsim k^{2}$ where $k$ is a running
scale. This is achieved by introducing a regulator that, either
sharply or smoothly, suppresses the contribution of modes with
$q^{2}\lesssim k^{2}$.  The FRG describes the evolution of this action
as the cut-off scale $k$ is lowered.  As  $k$
approaches zero, all fluctuations are included and thus the full
effective action is recovered.

The FRG method has already been applied to few-body problems in a
number of papers \cite{Di,Mor,Sch,KBW1,BKW}. In Ref.~\cite{Di}, it was
exploited to derive the Skornyakov--Ter-Martirosyan equation for three-body
systems
\cite{STM}.  The Efimov effect \cite{Efi} in bosonic and fermionic
systems was addressed in Ref.~\cite{Mor}. An important extension of
the approach was developed in Ref.~\cite{Sch} to treat the four-body
problem in the presence of the three-body Efimov effect, and the
process of dimer-dimer scattering was studied in detail in
Ref.~\cite{BKW}.  This shows one of the key strengths of the FRG
method, which is its universality. With relatively minor changes it
can be adopted to study a large variety of problems in
particle, nuclear and condensed matter physics. Some representative
examples can be found in Refs.~\cite{BKMW,Gie,DGPW}.

Up to now, most applications of the FRG have been to particle physics and 
condensed matter systems. In the present work we apply it to systems of up 
to four nucleons, laying the groundwork for extensions to larger numbers of 
nucleons and to nuclear matter. 
One important aspect of our implementation of the FRG is the introduction
of bosonic fields to describe interacting pairs of nucleons. This will be 
very useful for future work since it will allow us to link the scattering 
of nucleons in vacuum to pairing in systems of many nucleons, as in 
Ref.~\cite{BKMW}. This study of few-nucleon systems allows us to develop 
some of the tools and determine the parameters that will be needed in
such work. In this context, it is helpful to make use of Wigner's SU(4) 
``supermultiplet" symmetry, which is known to be a very good
approximate symmetry for light nuclei \cite{Wig},
 and even for some two-nucleon processes, provided the
  momenta involved are large compared with the inverse scattering
  lengths \cite{Mehen}.

The effective action is the Legendre transform of the
  logarithm of the partition function, $\Gamma[\phi_c]=W[J]-J\cdot
  \phi_c$, where $e^W$ is the partition function in the presence of an
  external source $J$ \cite{wein96}. The functional $\Gamma$ is the
  generator of the one-particle-irreducible Green's functions, and
  it reduces to the usual effective potential for homogeneous
  systems. One of the reasons to work with $\Gamma$ rather than $W$ is
  that we can introduce a renormalisation group flow to determine
  $\Gamma$ \cite{Wet}. A running version of the effective action is
  defined by introducing an artificial gap in the energy spectrum for
  the fields which depends on a momentum scale $k$.  Thus we define a
  different effective action for each $k$ by integrating over
  components of the fields with momenta $q>k$ only. The RG
  trajectory then interpolates between the classical action of the
  underlying field theory (at large $k$), and the full effective
  action (at zero $k$) \cite{BTW02}.  
The intermediate actions along the trajectory are not
    physically meaningful.

The flow equation for the effective action is a functional differential equation of the form
\begin{equation}
\partial_{k}\Gamma=-\frac{i}{2}\,\STr\left[(\partial_{k}R)\,(\Gamma^{(2)}-R)^{-1}\right].\label{eq:evolution}\end{equation}
where $\Gamma^{(2)}$ denotes the second functional derivative taken
with respect to the fields entering the action, and $R$ is a matrix of
cut-off functions. These functions act as regulators, suppressing the
contributions of fluctuations with momenta below the running scale
$k$, and thus driving the evolution of the system as $k$ is lowered.
The structure of the evolution equation is rather straightforward, the
main complexity is in the operation $\STr$, which is the supertrace,
which is taken over both energy-momentum variables and internal
indices. It is needed since we consider a mixed system of fermions
(nucleons) and bosonic dimers in this work.

The main advantage of this version of the RG is that the
  right-hand-side of Eq.~(\ref{eq:evolution}) involves a single
  integral over energy and momentum. It thus has the form of a
  one-loop integral, where $(\Gamma^{(2)}-R)^{-1}$ can be thought of
  as the matrix of single-particle propagators dressed by all
  fluctuations above the scale $k$. 
Since Eq.\ (\ref{eq:evolution}) is exact and therefore nonperturbative,
questions of diagrammatic
  expansion do not arise in this framework, in contrast to other
  versions of the RG and many traditional approaches to many-body
  physics. Nonetheless, despite its simple structure, this equation
  describes the running of a complicated object -- $\Gamma$ is in
  general a non-local functional of all the fields describing our
  system.

Since exact solution of a functional differential equation is usually not possible,
in practical applications, the REA is usually truncated to a finite
number of local terms. The choice of these is guided by the relevant
physics in the system we wish to describe, and by insights from
effective field theories. This truncation reduces the full functional
differential equation (\ref{eq:evolution}) to a set of coupled ODE's
for the running coupling constants and renormalisation factors
multiplying those terms in the action. These equations are
straightforward to solve 
nonperturbatively 
using standard methods.
It is the choice of the physics we want to describe that
  determines the form of the expansion of the REA.

The use of local interactions means that the FRG method has close
links to approaches based on effective field theories
\cite{Bea,BvK}. In some cases, for example the action used to study
the Efimov effect in three-boson systems, the terms included are just
the leading terms of the corresponding effective field theory. In
other cases, such as the applications to dense matter, strict power
counting arguments cannot be used and we must rely on the need to
describe emergent aspects of the physics that are known to be
important, such as superfluidity.

In the limit where $k$ is much larger that all the physical
  scales in our system, $k$ becomes the only important scale. The
  effects of the physical scales, for example the scattering lengths,
  become negligible in this regime and the evolution equations then
  show scaling behaviour with $k$. This behaviour is governed by the
  unitary limit, where the inverse two-body scattering lengths
  vanish. In this limit, all of the evolution equations collapse to
  simple universal forms for large enough values of the cut-off scale
  $k$.  Moreover, since all the inverse scattering lengths are
  negligible, these equations have SU(4)-symmetric forms.  These
universal equations can then be used to obtain boundary conditions for
the evolution equations away from the unitary limit. We present the
detailed form of these equations in the Appendix, where we show that they
form two decoupled sets of equations.

The set of equations that describes three or four particles
  in spatially symmetric states displays the Efimov effect
  \cite{Efi}. In these channels, scale invariance is broken by the
  appearance of an infinite tower of three-body bound states, with
  subsequent energies in a constant ratio. A single three-body initial
  condition needs to be specified to fix the energies of all these
  states. 

In the main body of the paper we derive a general set of equations for
the evolution of the REA, without assuming SU(4) symmetry.
Nonetheless, this symmetry plays a pivotal role in our
  approach. As just mentioned, the initial conditions are imposed at a
  scale where SU(4) is a very good symmetry. Initially the evolution
  remains in this regime, and it is only when it reaches scales $k\sim
  1/a$ that SU(4)-breaking effects can become large. In the two body
  sector, the symmetry is lost in the physical limit ($k\rightarrow
  0$) for quantities like the scattering lengths themselves. Perhaps
  surprisingly, SU(4) remains a good approximate symmetry for three-
  and four-body scattering at threshold. Working with the same
  combinations of couplings that decouple in the SU(4) limit, we find
  that the mixing remains small, except in narrow regions of the
  three-body parameter that controls the positions of the Efimov
  states.

One of these sets describes the channels where the nucleons are in states 
of mixed spatial symmetry and it has the same form as that
for a system of fermions with spin degrees of freedom only, as derived 
in Ref.~\cite{BKW}. The other set describes the spatially symmetric channels 
and, after a suitable redefinition of coupling constants, it matches the
evolution equations for a system of interacting bosons in Ref.~\cite{Mor}.
Despite the similar forms of the two sets of equations, the differences in
their numerical coefficients have profound physical consequences. In
particular, the equations for spatially symmetric systems leads to the 
Efimov effect \cite{Efi} and limit-cycle behaviour of the 3-body
couplings \cite{BHvK}.  In contrast these effects do not occur for 
few-fermion systems without spatial symmetry.

In the next section we set out the form of the REA we use in our
applications of the FRG to few-nucleon systems, and in Sec.~III we
obtain the evolution equations for the running parameters in that
action. Then, in Section IV, we apply these to analyse scattering in
three- and four-nucleon systems for realistic $NN$ scattering
lengths. We compare our results to those in the SU(4) limit, details
of which can be found in Appendix A.  Finally, some technical details
of the link between scattering lengths and REA are discussed in Appendix B.

\section{Running effective action}

Since our ultimate goal is to study pairing in dense nuclear matter,
we must isolate the relevant degrees of freedom. 
As is well known, the nuclear force is just
strong enough  to generate a bound state, the deuteron, in one of the $S$-wave channels.
Its isospin-1 analogue is only just unbound
and appears as a virtual state very close to threshold. We thus find
it useful to introduce boson fields to describe the lowest two-body
states in both these channels. If one imagines starting from an
effective action with a local two-body interaction between the
fermions, the boson fields can be introduced using a
Hubbard-Stratonovitch transformation, which is exact in this case. We
use the notation $\vvc$ for the (spin-triplet) vector-isoscalar boson
field, corresponding to the deuteron, and $\ivc$ for the
(spin-singlet) scalar-isovector one. These we refer to collectively as
{}``dimers'', and label by a capital $\boson$. We reserve the lower-case label
$d$ for a deuteron, and $\fermion$ for a generic nucleon. Finally,
when specifying quantum numbers of particular channels, we use the
$SO(4)$ notation $(S,I)$, where $S$ and $I$ denote the resultant spin
and isospin quantum numbers, respectively. In the present work, we
assume that isospin is a good symmetry. 
The use of such dibaryon fields was first suggested in Ref.\ \cite{Kaplan} in the context of effective field theories.

The key approximation we make is to truncate the full REA to one 
with only a finite number of parameters, so that the functional differential 
equation can be replaced by a set of ordinary differential equations. 
No further approximations are made, which means that the solution
of the resulting equations is nonperturbative in all their couplings. 
In the 
present case, we introduce a set of local contact interactions in the channels
of interest, as well as kinetic terms for the dimer fields.
In principle we could also have added effective ranges for
  the interactions using the forms used in effective field theory; in
  this first application we have not included such effects.
The form for the REA we use is
\begin{eqnarray}
\Gamma & = & \int d^{4}p\,\psi_{m_{\vsum}m_{\isum}}^{\dagger}(p_{0},\vec{p})\left(p_{0}-\frac{p^{2}}{2M}+i\epsilon\right)\psi_{m_{\vsum}m_{\isum}}(p_{0},\vec{p})\nonumber \\
 &  & +\int d^{4}p\,\vb_{i}^{\dagger}(p_{0},\vec{p})\left(Z_{\phi,\vb}p_{0}-Z_{m,\vb}\frac{p^{2}}{4M}-u_{1,\vb}+i\epsilon\right)\vb_{i}(p_{0},\vec{p})\nonumber \\
 &  & +\int d^{4}p\,\ib_{a}^{\dagger}(p_{0},\vec{p})\left(Z_{\phi,\ib}p_{0}-Z_{m,\ib}\frac{p^{2}}{4M}-u_{1,\ib}+i\epsilon\right)\ib_{a}(p_{0},\vec{p})\nonumber \\
 &  & +\Gamma_{2}+\Gamma_{3}+\Gamma_{4}.
\end{eqnarray}
Since the action at the start of the evolution is obtained by bosonising
a purely fermionic one, the boson fields should be non-propagating
auxiliary fields for $k\rightarrow\infty$. Their wave function and
kinetic-mass renormalisation factors ($Z_{\phi,i}$ and $Z_{m,i}$)
should tend to zero in this limit.  When we run the action as we lower
the cut-off scale $k$, these fields become dynamical, and the renormalisation
factors grow.

The bosonisation introduces self-energies for the boson fields ($u_{1,i}$) 
and couplings of the dimers to pairs of nucleons in the $(1,0)$ and $(0,1)$ 
two-body channels. These couplings are described by the term,\begin{eqnarray}
\Gamma_{2} & = & \int\delta^{(4)}(p_{1}+p_{2}-p_{3})d^{4}p_{1}d^{4}p_{2}d^{4}p_{3}\mathcal{V}_{2}\\
\mathcal{V}_{2} & = & \frac{1}{2\sqrt{2}}g_{\vsum}\left[\vec{\vb}^{\dagger}(p_{30},\vec{p}_{3})\cdot\left[\psi(p_{10},\vec{p}_{1})\psi^{C}(p_{20},\vec{p}_{2})\right]^{(1,0)}+\text{h.c.}\right]\nonumber \\
 &  & +\frac{1}{2\sqrt{2}}g_{\isum}\left[\vec{\ib}^{\dagger}(p_{30},\vec{p}_{3})\cdot\left[\psi(p_{10},\vec{p}_{1})\psi^{C}(p_{20},\vec{p}_{2})\right]^{(0,1)}+\text{h.c.}\right],\label{eq:V2DN}\end{eqnarray}
where 
\begin{equation}
\psi_{m_{\vsum_2}m_{\isum_2}}^{C}=\tau_{m_{\isum_1}m_{\isum_2}}^{2}\sigma_{m_{\vsum_1}m_{\vsum_2}}^{2}\psi_{m_{\vsum_1}m_{\isum_1}}.
\end{equation}
Since $\vvc^{\dagger}$ is a $(1,0)$ tensor and $\ivc^{\dagger}$ a $(0,1)$ one,
the scalar products in Eq.~(\ref{eq:V2DN}) should be understood as being taken
in either spin or isospin space, as appropriate. The values of the 
coupling constants $g_{i}$ 
depend on a choice of scale in the dimer fields when they are introduced by
the Hubbard-Stratonovitch transformation. Thus we must find that the
values of the $g_{i}$ do not appear separately in physical quantities, 
but only in combinations such as $g_i^2/u_{1,i}$. An illustration
of this is provided by the discussion of the evolution of the $u_{1,i}$
in the next section. The unobservable coupling constants $g_{i}$ do not run 
in vacuum and, for simplicity, we choose their values to be equal,
\begin{equation}
g_{\vsum,\isum}=g.
\end{equation}

To describe the three-nucleon channels, we introduce a set of local dimer-nucleon
interactions, described by the term 
\begin{eqnarray}
\Gamma_{3} & = & -\int\delta^{(4)}(p_{1}+p_{3}-p_{2}-p_{4})d^{4}p_{1}d^{4}p_{2}d^{4}p_{3}d^{4}p_{4}\mathcal{V}_{3,\boson\fermion}\\
\mathcal{V}_{3,\boson\fermion} & = & \sum_{i,j=\vsum,\isum}\lambdahh_{ij}\left[i^{\dagger}(p_{30},\vec{p}_{3})\psi^{\dagger}(p_{10},\vec{p}_{1})\right]^{(1/2,1/2)}\cdot\left[\psi(p_{20},\vec{p}_{2})j(p_{40},\vec{p}_{4})\right]^{(1/2,1/2)}\nonumber \\
 &  & \quad+\lambdath_{\vsum\vsum}\left[\vb^{\dagger}(p_{30},\vec{p}_{3})\psi^{\dagger}(p_{10},\vec{p}_{1})\right]^{(3/2,1/2)}\cdot\left[\psi(p_{20},\vec{p}_{2})\vb(p_{40},\vec{p}_{4})\right]^{(3/2,1/2)}\nonumber \\
 &  & \quad+\lambdaht_{\isum\isum}\left[\ib^{\dagger}(p_{30},\vec{p}_{3})\psi^{\dagger}(p_{10},\vec{p}_{1})\right]^{(1/2,3/2)}\cdot\left[\psi(p_{20},\vec{p}_{2})\ib(p_{40},\vec{p}_{4})\right]^{(1/2,3/2)},\end{eqnarray}
where $\lambdahh_{\vb\ib}=\lambdahh_{\ib\vb}$.
These have been expressed in terms of interactions in the doublet-doublet channel,
with spin-isospin quantum numbers $(1/2,1/2)$, and the quartet-doublet channels, 
with quantum numbers $(3/2,1/2)$ or $(1/2,3/2)$. The $(1/2,1/2)$ channel has the
quantum numbers of the ground states of $^3$H and $^3$He.

Finally, we introduce one class of four-nucleon interactions, represented by 
local two-body dimer-dimer interactions and described by the term
\begin{eqnarray}
\Gamma_{2,\boson\boson} & = & -\int\delta^{(4)}(p_{1}+p_{3}-p_{2}-p_{4})d^{4}p_{1}d^{4}p_{2}d^{4}p_{3}d^{4}p_{4}\mathcal{V}_{2,\boson\boson},\\
\mathcal{V}_{2,\boson\boson} & = & \frac{1}{2}\sum_{i=\vsum,\isum}u_{2,i}\biggl(\vec{i}^{\dagger}(p_{10},\vec{p}_{1})\cdot\vec{i}(p_{20},\vec{p}_{2})\,\vec{i}^{\dagger}(p_{30},\vec{p}_{3})\cdot\vec{i}(p_{40},\vec{p}_{4})\nonumber \\
 &  & \qquad-\frac{1}{3}\vec{i}^{\dagger}(p_{10},\vec{p}_{1})\cdot\vec{i}^{\dagger}(p_{30},\vec{p}_{3})\,\vec{i}(p_{20},\vec{p}_{2})\cdot\vec{i}(p_{40},\vec{p}_{4})\biggr)\nonumber \\
 &  & +u_{2,\vsum\isum}\biggl(\vec{\vsum}^{\dagger}(p_{10},\vec{p}_{1})\cdot\vec{\vsum}(p_{20},\vec{p}_{2})\,\vec{\isum}^{\dagger}(p_{30},\vec{p}_{3})\cdot\vec{\isum}(p_{40},\vec{p}_{4})\biggr)\nonumber \\
 &  & +\frac{1}{12}\sum_{ij=\vsum,\isum}\bar{u}{}_{2,ij}\vec{i}^{\dagger}(p_{10},\vec{p}_{1})\cdot\vec{i}^{\dagger}(p_{30},\vec{p}_{3})\,\vec{j}(p_{20},\vec{p}_{2})\cdot\vec{j}(p_{40},\vec{p}_{4}),\end{eqnarray}
where $\bar{u}{}_{2,\vb\ib}=\bar{u}{}_{2,\ib\vb}$.
The $u_{2,i}$ terms act in the $(2,0)$
and $(0,2)$ two-dimer (four-nucleon) channels, the $u_{2,\vsum\isum}$
terms act in the $(1,1)$ channel, and the $\bar{u}_{2}$ terms act
in the $(0,0)$ channel. Only the last of these has the quantum
numbers of the ground state of the $\alpha$ particle.

The behaviour for large $k$ is particularly simple.  In this region,
the regulator provides the only scale and every coupling varies as a power
of $k$. The effects of SU(4)-breaking also become negligible here and
so the evolution equations reduce to the simpler, SU(4)-symmetric
forms discussed in Appendix A.  This allows us to determine the
initial conditions on the evolution here and these then provide the
physical input into our approach.

Most of the parameters are ``irrelevant" in the technical sense that the 
low-energy results from our action in the physical limit ($k\rightarrow 0$)
do not depend strongly on the precise values of the initial conditions imposed 
at very large $k$. However, there are three exceptions and these provide
the physical input into our approach. Two of them are the self-energies, $u_{1,i}$,
which are ``relevant" parameters in RG language, as discussed in Ref.~\cite{Bir}.
Their values at $k=0$ determine the strength of the interaction between the fermions
and can be related either to the $NN$ scattering lengths, $a_\vb$ and $a_\ib$,
or, in the channel with the bound state, to the deuteron binding energy,
\begin{equation}
\ed=-\frac{1}{Ma_{\vb}^{2}}.
\label{eq:ed}
\end{equation}

The third parameter is associated with the spatially-symmetric channel of the 
three-nucleon system. This is the channel that displays the Efimov effect 
\cite{Efi} and as a result the evolution of its coupling constant for it 
has a limit-cycle behaviour at large-$k$. Just as in the analogous EFT 
treatment \cite{BHvK}, one piece of three-body data is needed to fix the 
starting point on this cycle. Here we choose the nucleon-deuteron 
scattering length in the spin-doublet channel, and explore how four-nucleon
observables are related to it.

\section{Evolution equations}

To derive the evolution equations for the coupling constants in our action, 
we substitute the parametrisation given in the previous section into the 
right-hand side of Eq.~(\ref{eq:evolution}). The inverse of the second 
derivative of $\Gamma$ forms a scale-dependent propagator, and the driving 
terms for the coupling constants all have structure of one-loop integrals, 
with differing numbers of external fields. In general these include 
expressions with powers of fields or derivatives beyond those contained 
in our truncated action. We therefore need to expand the loop integrals
in powers of the fields and non-localities (i.e., in energies and momenta)
and pick out those terms that match the structures listed in the previous 
section. The scale-dependent coefficient in front of each structure 
then gives us the driving term in the differential equation for the 
corresponding coupling constant.

As well as the choice of truncation, we also need to make a choice of the
point around which we expand our energies. 
Since we concentrate on the physics at the dimer threshold,
  we choose our expansion points for this state; other choices are
  possible, but suffer from numerical complications arising from
  linking the flows both above and below threshold.
For the dimers, we choose the
energy of the lowest two-body bound state, $\ed$. In the four-nucleon sector,
for example, this means that we expand around the deuteron-deuteron threshold.
For the nucleons we expand around one half of this binding energy. 
Thus in the nuclear-dimer channels of the the three-nucleon system we
expand about an energy $\ed/2$ below the the scattering threshold.
This 
procedure can be justified rigorously in the limit of exact SU(4) symmetry, as
discussed further below, but it is only approximate in the general case. It 
could thus lead to a slower convergence than expected of the expansion.

Finally, we need to choose the forms for the cut-off functions in the regulators 
for the nucleons and dimers. Here we take the form suggested by Litim
\cite{Lit} for both, as it is well-suited for partially analytic calculations. 
It is also optimised for actions truncated to purely local (energy-independent)
interactions, as also discussed in more detail by Pawlowski \cite{Paw}.
The regulators are thus given by 
\begin{align}
R_{\fermion}(q;k) & =\frac{k^{2}-q^{2}}{2M}\theta(k-q),\label{eq:Rferm}\\
R_{\boson}(q;k) & =Z_{\phi}(k)\frac{k^{2}-q^{2}}{4M}\theta(k-q).\label{eq:Rbos}
\end{align}
Note that we have assumed that the two $Z_{\phi}$'s are equal; this can be shown to hold 
in the expansion scheme we use here, see Eq.~(\ref{eq:Zphi}) below.
The inclusion of the wave-function renormalisation factor in the definition
of $R_{\boson}$ has the advantage of allowing us to scale out the parameters
$g$, $M$ and $a_{\vb}$, leaving much simpler, dimensionless expressions.
For example, a generic three-body coupling $\lambda$ has  a natural scaling 
of the form
\begin{equation}
\Lambda(\kappa)=\frac{1}{a_{\vb}^{2}Mg^{2}}\lambda(k).
\label{eq:Lamdefn}
\end{equation}
Analogously a generic four-body coupling $u_2$ can be scaled as 
\begin{equation}
U_{2}(\kappa)=g^{-4}M^{-3}a_{\vb}^{-3}u_{2}(k).
\label{eq:U2defn}
\end{equation}

\subsection{One-boson terms}

The evolution equations for $u_{1,i}$, $Z_{\phi,i}$ and $Z_{m,i}$
are relatively straightforward in vacuum. 
Their forms are the same as in the simpler case studied in Ref.~\cite{BKW};
For example, both the $u_{1,i}$ satisfy the differential equation
\begin{equation}
\partial_{k}u_{1,i}(k)=\frac{g^{2}}{2}\frac{1}{\left(2\pi\right)^{3}}
\int d^{3}q\frac{\partial_{k}R_{{\fermion}}(q;k)}{\left(\frac{q^{2}}{2M}
+R_{{\fermion}}(q;k)-\ed/2\right)^{2}},\label{eq:dku1}
\end{equation}
where $\ed$ appears in the denominator as a result of our choice of expansion 
point and we have already performed the integral over the virtual energy $q^0$. 
This equation is an exact differential and can be integrated directly. Its
solution grows linearly with $k$ as $k\rightarrow\infty$, reflecting
the relevant nature of the parameters $u_{1,i}$.

The difference between the two channels is encoded only in different boundary
conditions imposed on the solutions to this equation. In the case of the $\vb$ 
channel, there is a bound state and we have chosen to expand in powers of the 
energy, $p_0$, relative to this state. The dimer propagator in this channel 
should therefore have a pole at $p_0=0$ in the physical limit, which leads to
the condition 
\begin{equation}
u_{1,\vb}(k=0)=0.\label{eq:u1sBC}
\end{equation}
The solution to Eq.~(\ref{eq:dku1}) that satisfies this can be written in 
the form
\begin{equation}
u_{1,\vb}(k)  =  -\frac{g^{2}M}{4\pi a_{\vb}}-\frac{g^{2}}{2}\frac{1}{\left(2\pi\right)^{3}}\int d^{3}q\left[\frac{1}{\frac{q^{2}}{2M}+R_{{\fermion}}(q)-\ed/2}-\frac{1}{\frac{q^{2}}{2M}}\right].
\label{eq:u1s1}
\end{equation}
For Litim's regulator, Eq.\ (\ref{eq:Rferm}), this gives
\begin{equation}
u_{1,\vb}(k) =  \frac{g^{2}M}{4\pi a_{\vb}}\left(\frac{4}{3\pi}\kappa+\frac{2}{3\pi}\frac{\kappa}{\left(\kappa^{2}+1\right)}+\frac{2}{\pi}\cot^{-1}(\kappa)-1\right),
\label{eq:u1s}
\end{equation}
where we have used Eq.~(\ref{eq:ed}) for the deuteron binding energy and we have
introduced the dimensionless variable $\kappa=k a_{\vb}$.

The fact that SU(4) is not an exact symmetry means that the two scattering lengths
are not equal. In particular, the $\ib$ channel has no bound state and hence 
a negative scattering length, $a_\ib$. If this were the only channel we were 
interested in, we could just expand energies around zero and the appropriate 
boundary condition would be 
\begin{equation}
u_{1,\ib}(k=0)=-\frac{g^{2}M}{4\pi a_\ib},
\label{eq:u1sBCsc}\end{equation}
which is obtained by replacing $a_\vb$ by $a_\ib$ and setting $\ed$ to
zero in Eq.\ (\ref{eq:u1s1}).  However we cannot use this here since we need to treat both
channels simultaneously and, as discussed above, we haven taken the
deuteron energy as our expansion point. (A similar issue would arise
even if the $\ib$ channel also had a bound state but this was
shallower than the deuteron.) The appropriate boundary condition can be obtained by replacing $a_\vb$
by $a_\ib$ in the first part of the right-hand side of
Eq.~(\ref{eq:u1s1}), while keeping $\ed$ in the second term.
 Introducing the
dimensionless parameter $\alpha$ ($\leq 1$) by
\begin{equation}
1/a_{\ib}=\alpha/a_{\vb},
\end{equation}
we can write the solution in the form
\begin{equation}
u_{1,\ib}(k)=u_{1,\vb}(k)+\frac{g^{2}M}{4\pi a_{\vb}}\left(1-\alpha\right).
\end{equation}
The parameter $1-\alpha$ provides a measure for the symmetry
  breaking in the two-body channels. We treat this breaking
  nonperturbatively in the following calculations.

The wave function renormalisation factors $Z_{\phi,i}$ both satisfy the same 
equation and we can impose the same boundary condition that they vanish as 
$k\rightarrow\infty$ on them. They then have the same form in both channels,
\begin{eqnarray}
Z_{\phi,i}(k) & = & \frac{1}{4}\frac{1}{\left(2\pi\right)^{3}}\int d^{3}q\frac{1}{\left[\frac{q^{2}}{2M}+R_{{\fermion}}(q)-\ed/2\right]^{2}}\nonumber \\
 & = & \frac{a_{\vb}g^{2}M^{2}}{8\pi}\left(\frac{2\kappa\left(5\kappa^{2}+3\right)}{3\pi\left(\kappa^{2}+1\right)^{2}}+\frac{2}{\pi}\cot^{-1}(\kappa)\right).
\label{eq:Zphi}
\end{eqnarray}

Litim's cut-off \cite{Lit} respects Galilean invariance to the order we work here 
(see Ref.~\cite{Bir} for more details) and we find that the mass renormalisation
factors are the same as those multiplying the energy,
\begin{equation}
Z_{m,i}=Z_{\phi,i}.
\end{equation}

To simplify the expressions for the loop integrals in channels with more than
two particles, we introduce a compact notation for the inverse propagators,
\begin{eqnarray}
E_{{\fermion}R}(q) & = & \frac{q^{2}}{2M}-\ed/2+R_{{\fermion}}(q),\label{eq:EFR}\\
E_{{\boson}R,i}(q) & = & Z_{\phi}\left[\frac{q^{2}}{4M}+u_{1,i}(q)/Z_{\phi}-\ed+R_{{\boson}}(q)/Z_{\phi}\right],\ \label{eq:EBR}\end{eqnarray}
where we have suppressed the implicit $k$ dependence of the running quantities
in these expressions.

\subsection{Three-body couplings}

The evolution equations for the three-body (nucleon-dimer) couplings decouple 
into a set of four for the $(1/2,1/2)$ channel and two separate equations for the 
$(3/2,1/2)$ and $(1/2,3/2)$ channels. They have the forms
\begin{eqnarray}
\partial_{k}\lambdahh_{\vsum\vsum} & = & \lambdahh_{\isum\vsum}\lambdahh_{\vsum\isum}I_{1,\ib}+\left(\lambdahh_{\vsum\vsum}\right)^{2}I_{1,\vb}\nonumber \\
 &  & +g^{2}\left(\frac{3}{4}\left[\lambdahh_{\isum\vsum}+\lambdahh_{\vsum\isum}\right]I_{2,\vb}-\frac{1}{2}\lambdahh_{\vsum\vsum}I_{2,\ib}\right)\nonumber \\
& & +\frac{1}{16}g^{4}\left(I_{3,\vb}+9I_{3,\ib}\right),\label{eq:lamvv}\\
\partial_{k}\lambdahh_{\vsum\isum} & = & \lambdahh_{\vsum\vsum}\lambdahh_{\vsum\isum}I_{1,\vb}+\lambdahh_{\vsum\isum}\lambdahh_{\isum\isum}I_{1,\ib}\nonumber \\
 &  & +\frac{1}{4}g^{2}\left(\left[3\lambdahh_{\vsum\vsum}-\lambdahh_{\vsum\isum}\right]I_{2}^{\vsum}+\left[3\lambdahh_{\isum\isum}-\lambdahh_{\vsum\isum}\right]I_{2}^{\isum}\right)\nonumber \\
 &  & -\frac{3}{16}g^{4}\left(I_{3,\vb}+I_{3,\ib}\right),\label{eq:lamvi}\\
\partial_{k}\lambdath_{\vsum\vsum} & = & \left(\lambdath_{\vsum\vsum}\right)^{2}I_{1,\vb}+g^{2}\lambdath_{\vsum\vsum}I_{2,\vb}+\frac{1}{4}g^{4}I_{3,\vb},\end{eqnarray}
where we have displayed only half the equations; the others can obtained
by appropriate interchanges of spin and isospin labels, in particular 
$\vsum\leftrightarrow\isum$. We have also defined a short hand for the
$k$-dependant loop integrals
\begin{eqnarray}
I_{1,i} & = & \frac{1}{(2\pi)^{3}}\int d^{3}q\,\frac{\partial_{k}R_{{\boson}}+Z_{\phi}\partial_{k}R_{{\fermion}}}{\left(E_{{\boson}R,i}(q)+Z_{\phi}E_{{\fermion}R}(q)\right)^{2}},\\
I_{2,i} & = & \frac{1}{(2\pi)^{3}}\int d^{3}q\,\frac{E_{{\fermion}R}(q)\partial_{k}R_{{\boson}}+\left(E_{{\boson}R,i}+2Z_{\phi}E_{{\fermion}R}\right)\partial_{k}R_{{\fermion}}}{E_{{\fermion}R}(q)^{2}\left(E_{{\boson}R,i}(q)+Z_{\phi}E_{{\fermion}R}(q)\right)^{2}},\\
I_{3,i} & = & \frac{1}{(2\pi)^{3}}\int d^{3}q\,\frac{E_{{\fermion}R}(q)\partial_{k}R_{{\boson}}+\left(2E_{{\boson}R,i}+3Z_{\phi}E_{{\fermion}R}\right)\partial_{k}R_{{\fermion}}}{E_{{\fermion}R}(q)^{3}\left(E_{{\boson}R,i}(q)+Z_{\phi}E_{{\fermion}R}(q)\right)^{2}},\end{eqnarray}
which can be evaluated in closed form for our choice of cut-off functions,
Eqs. (\ref{eq:EFR},\ref{eq:EBR}).

\subsection{Dimer-dimer couplings}

The evolution equations for the four-body (dimer-dimer) couplings
$u_{2}$ or $\bar{u}_{2}$ also decouple, into a set of four for the
$(0,0)$ channel and three separate equations for the $(2,0)$, $(0,2)$
and $(1,1)$ channels. They have the forms (where again we have
displayed only half the equations),
\begin{eqnarray}
\partial_{k}u_{2,\vsum} & = & \frac{1}{2}u_{2,\vsum}^{2}K_{1,\vb}-2g^{2}
\lambdath_{\vsum\vsum}K_{2}-\frac{3}{4}g^{4}K_{3},\label{eq:u2tevo}\\
\partial_{k}u_{2,\vsum\isum} & = & \frac{1}{2}u_{2,\vsum\isum}^{2}K_{1,\vb\ib}-g^{2}\frac{1}{3}\left(2\sum_{i=\vb,\ib}\lambdath_{ii}+\sum_{ij=\vb,\ib}\lambdahh_{ij}\right)K_{2}-\frac{3}{4}g^{4}K_{3},\\
\partial_{k}\bar{u}_{2\vsum\vsum} & = & \frac{1}{4}\bar{u}_{2\vsum\vsum}^{2}K_{1,\vb}+\frac{1}{4}\bar{u}_{2\vsum\isum}\bar{u}_{2\isum\vsum}K_{1,\ib}-12g^{2}\lambdahh_{\vsum\vsum}K_{2}+\frac{3}{4}g^{4}K_{3},\\
\partial_{k}\bar{u}_{2\vsum\isum} & = & \frac{1}{4}\bar{u}_{2\vsum\vsum}\bar{u}_{2\isum\vsum}K_{1,\vb}+\frac{1}{4}\bar{u}_{2\isum\isum}\bar{u}_{2\vsum\isum}K_{1,\ib}-12g^{2}\lambdahh_{\isum\vsum}K_{2}-\frac{9}{4}g^{4}K_{3}.\label{eq:u2btsevo}\end{eqnarray}
Note that each equation contains only a limited subset of the thee-body couplings, as a result of the constraints of angular momentum and isospin. The loop integrals used 
here are 
\begin{eqnarray}
K_{1,i} & = & \frac{1}{(2\pi)^{3}}\int d^{3}q\frac{\partial_{k}R_{{\boson}}}{Z_{\phi}E_{{\boson}R,i}(q)^{2}},\\
K_{1,\vsum\isum} & = & \frac{1}{(2\pi)^{3}}\int d^{3}q\frac{4\partial_{k}R_{{\boson}}}{Z_{\phi}\left(E_{{\boson}R,\vb}(q)+E_{{\boson}R,\ib}(q)\right)^{2}},\\
K_{2} & = & \frac{1}{(2\pi)^{3}}\int d^{3}q\frac{\partial_{k}R_{{\fermion}}}{E_{{\fermion}R}(q)^{3}},\\
K_{3} & = & \frac{1}{(2\pi)^{3}}\int d^{3}q\frac{\partial_{k}R_{{\fermion}}}{E_{{\fermion}R}(q)^{4}}.\end{eqnarray}

\subsection{SU(4)-breaking form}

Since the 
inverse scattering lengths for nucleons are all small
  compared to the typical momentum scales of bound states of more than
  two nucleons, SU(4) is a good approximate symmetry in nuclear
  physics. Also, in the context of our RG treatment, much of the
  evolution takes place on momentum scales where SU(4) is a very good
  approximate symmetry. 
It is therefore
sensible to re-express our couplings in terms of ones which multiply 
SU(4)-symmetric combinations of fields, and ones which are generated by 
SU(4)-breaking in the two-body sector. This is also needed for analysing the 
behaviour of the couplings in the large-$k$ regime where we impose our initial
conditions, as discussed above. Using the same notation as in Appendix A and 
denoting the SU(4)-breaking couplings with a ``$\delta$'', we define, for example,
\begin{eqnarray}
\lambda & = & \frac{1}{2}
\left( \lambdahh_{\vsum\vsum}+\lambdahh_{\isum\isum}-2\lambdahh_{\isum\vsum}\right),
\label{eq:lamSU4}\\
\lambda^{\prime} & = & \frac{1}{2}
\left(  \lambdahh_{\vsum\vsum}+\lambdahh_{\isum\isum}+2\lambdahh_{\isum\vsum}\right),
\label{eq:lampSU4}\\
\delta\lambda & = & \lambdahh_{\vsum\vsum}-\lambdahh_{\isum\isum},
\label{eq:deltalamSU4}\\
\bar{u}_{2}^{\prime} & = & \frac{1}{2}
\left(
\bar{u}_{2\isum\isum}+\bar{u}_{2\vsum\vsum}+2\bar{u}_{2\vsum\isum}\right),\\
\bar{u}_{2} & = & \frac{1}{2}
\left(\bar{u}_{2\isum\isum}+\bar{u}_{2\vsum\vsum}
-2\bar{u}_{2\vsum\isum}\right),\\
\delta\bar{u}_{2} & = & \frac{1}{2}\left(\bar{u}_{2\vsum\vsum}-\bar{u}_{2\isum\isum}\right),
\end{eqnarray}
for the $(1/2,1/2)$ and $(0,0)$ channels. In a similar way we break the loop 
integrals up into, for example,
\begin{eqnarray}
I_{j} & = & \frac{1}{2}\left(I_{j,\isum}+I_{j,\vsum}\right),\\
\delta I_{j} & = & I_{j,\vsum}-I_{j,\isum}.
\end{eqnarray}

The equations we obtain are generalisations of the SU(4) symmetric
case discussed in Appendix A.  For the spatially symmetric channels
they become
 \begin{eqnarray}
\partial_{k}\lambda & = & I_{1}\lambda^{2}+g^{2}I_{2}\lambda+\frac{1}{4}g^{4}I_{3}\nonumber \\
 &  & +\frac{9}{4}I_{1}\delta\lambda^{2}+\delta\lambda\left[\frac{3}{2}\delta I_{1}\lambda+\frac{3}{4}g^{2}\delta I_{2}\right],\\
\partial_{k}\lambda^{\prime} & = & I_{1}\lambda^{\prime2}-2g^{2}I_{2}\lambda^{\prime}+g^{4}I_{3}\nonumber \\
 &  & +\frac{9}{4}I_{1}\delta\lambda^{2}+\delta\lambda\left[\frac{3}{2}\delta I_{1}\lambda^{\prime}-\frac{3}{2}g^{2}\delta I_{2}\right],\\
\partial_{k}\delta\lambda & = & \frac{3}{4}\delta I_{1}\delta\lambda^{2}+I_{1}\delta\lambda(\lambda+\lambda^{\prime})-\frac{1}{2}g^{2}I_{2}\delta\lambda\nonumber \\
 &  & +\frac{1}{3}\delta I_{1}\lambda\lambda^{\prime}+g^{2}\frac{1}{6}\delta I_{2}(\lambda^{\prime}-2\lambda)-\frac{1}{6}g^{4}\delta I_{3},\\
\partial_{k}\bar{u}_{2}^{\prime} & = & \frac{1}{2}K_{1}\left(\bar{u}_{2}^{\prime}\right)^{2}-2g^{2}K_{2}\lambda-\frac{3}{4}g^{4}K_{3}\nonumber \\
 &  & +\frac{1}{2}K_{1}\left(\delta\bar{u}_{2}\right)^{2}+\delta K_{1}\delta\bar{u}_{2}\bar{u}_{2}^{\prime},\\
\partial_{k}\bar{u}_{2} & = & \frac{1}{2}K_{1}\bar{u}_{2}^{2}-2g^{2}K_{2}\lambda^{\prime}+\frac{3}{2}g^{4}K_{3}\nonumber \\
 &  & +\frac{1}{2}K_{1}\left(\delta\bar{u}_{2}\right)^{2}+\delta K_{1}\delta\bar{u}_{2}\bar{u}_{2},\\
\partial_{k}\delta\bar{u}_{2} & = & \frac{1}{2}\delta K_{1}\left(\delta\bar{u}_{2}\right)^{2}-3g^{2}K_{2}\delta\lambda\nonumber \\
 &  & +\frac{1}{2}K_{1}\delta\bar{u}_{2}(\bar{u}_{2}^{\prime}+\bar{u}_{2})+\frac{1}{2}\delta K_{1}\bar{u}_{2}^{\prime}\bar{u}_{2}.\end{eqnarray}
Note that all SU(4)-breaking terms on the right-hand-sides are at least 
quadratic in symmetry breaking quantities. 
Even though we shall solve the  set of equations
(\ref{eq:lamvv}-\ref{eq:lamvi}) and
(\ref{eq:u2tevo}-\ref{eq:u2btsevo}), without imposing SU(4) symmetry,
the reorganisation of the potential in this section leads us to expect that
generically 
SU(4) is a rather good approximate symmetry in the three- and
four-body systems.

\section{Results}

\subsection{SU(4)-symmetric limit}

We first look at the case of exact supermultiplet symmetry, $a_{\ib}=a_{\vb}$.
The evolution equations in this limit are described in detail in Appendix A.
As explained there, we can reduce the problem to a limited set of
parameters. These include two dimensionless three-body coupling constants,
one of which (the coupling $\lambda^{\prime}$ in the (1/2,1/2) channel) 
exhibits the Efimov effect \cite{Efi}. 

This effect is a remarkable feature of any three-body system with a
attractive short-range interaction. In the unitary limit 
($a\rightarrow\infty$) such a system possesses an infinite number 
of there-body bound states with a geometric spectrum.
Even away from the unitary limit, these systems display universal 
features, such as relations between various three- and four-body 
observables \cite{Phi,Tjon}. In the framework of a renormalisation group, 
the evolution of the corresponding three-body force shows a limit-cycle 
behaviour \cite{BHvK}, as consequence of the periodic appearance of new 
Efimov states as the cut-off is lowered. Away from the unitary limit, the 
finite inverse scattering length provides an infrared cut-off on the Efimov 
behaviour. This leaves a unique shallowest bound state which, in the 
nuclear context, we interpret as the triton. As a result, the periodic
behaviour of the three-body coupling stops when the cut-off scale 
decreases to a value comparable with $1/a$ where it becomes almost 
independent of the running scale. More details of the running of 
$\lambda^{\prime}$ in the unitary limit can be found in Appendix A.

\begin{figure}
\begin{centering}
\includegraphics[width=6cm]{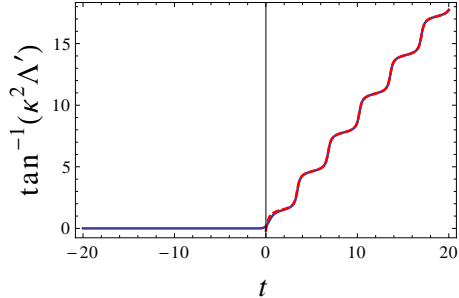}
\par\end{centering}
\caption{{(}Colour online{)} An example of the full evolution of $\Lambda^{\prime}$ (solid blue
line) compared to the asymptotic evolution (red dashed line). The
full evolution was started by integrating downwards from $t=20$,
with the same initial condition as chosen for the asymptotic solution. 
 \label{fig:lambdap}}
\end{figure}

We first compare the asymptotic results for $\lambda^{\prime}$, or
rather a rescaled version of this coupling, $\Lambda^{\prime}$, defined as in 
Eq.~(\ref{eq:Lamdefn}), to the numerical solution of the full equation.
This coupling diverges periodically in $t=\ln(\kappa)=\ln(ka_\vb)$,
reflecting the appearance of the geometrically spaced states of the Efimov 
effect. In Fig.~\ref{fig:lambdap} we plot the arctangent of this coupling 
(after a suitable rescaling), which remains finite. The linear growth seen 
for large $t$ is the signal of Efimov behaviour. The plot shows that the 
full solution follows the asymptotic form very closely down to
$t\simeq 0$. At this point, the scale $1/a_\vb$ becomes important and 
acts as a low-energy cut-off on the tower of Efimov states. 

If we add the appropriate $i\epsilon$ terms to impose causal boundary 
conditions on the propagators, we find that $\Lambda^{\prime}$ has 
an infinitesimal negative imaginary part. In our numerical treatment we take
advantage of this, by starting the evolution of $\Lambda^{\prime}$ at 
large $\kappa$ from the asymptotic solution,
\begin{equation}
\kappa^{2}\ensuremath{\Lambda^{\prime}}(t)=\frac{1}{28}\left(31-5\sqrt{535}\tan\left[\frac{1}{25}\left(5-\delta i-\sqrt{535}t\right)\right]\right),
\label{eq:initLam}
\end{equation}
with small a imaginary part $\delta$. As we evolve downward in $t$, this allows
our solution to bypass the singularities on the correct side and hence we are able 
to integrate the equations for the four-body couplings numerically. In this context we 
note that it is important not to take the finite imaginary part to be 
too small relative to the numerical precision used in the integration,
as otherwise errors are introduced by the integration regions close to
the singularities. Our results in the scaling region differ from those 
of Schmidt and Moroz \cite{Sch}, who took too small values for the 
imaginary part.

\begin{figure}
\begin{centering}
\includegraphics[clip,width=6cm]{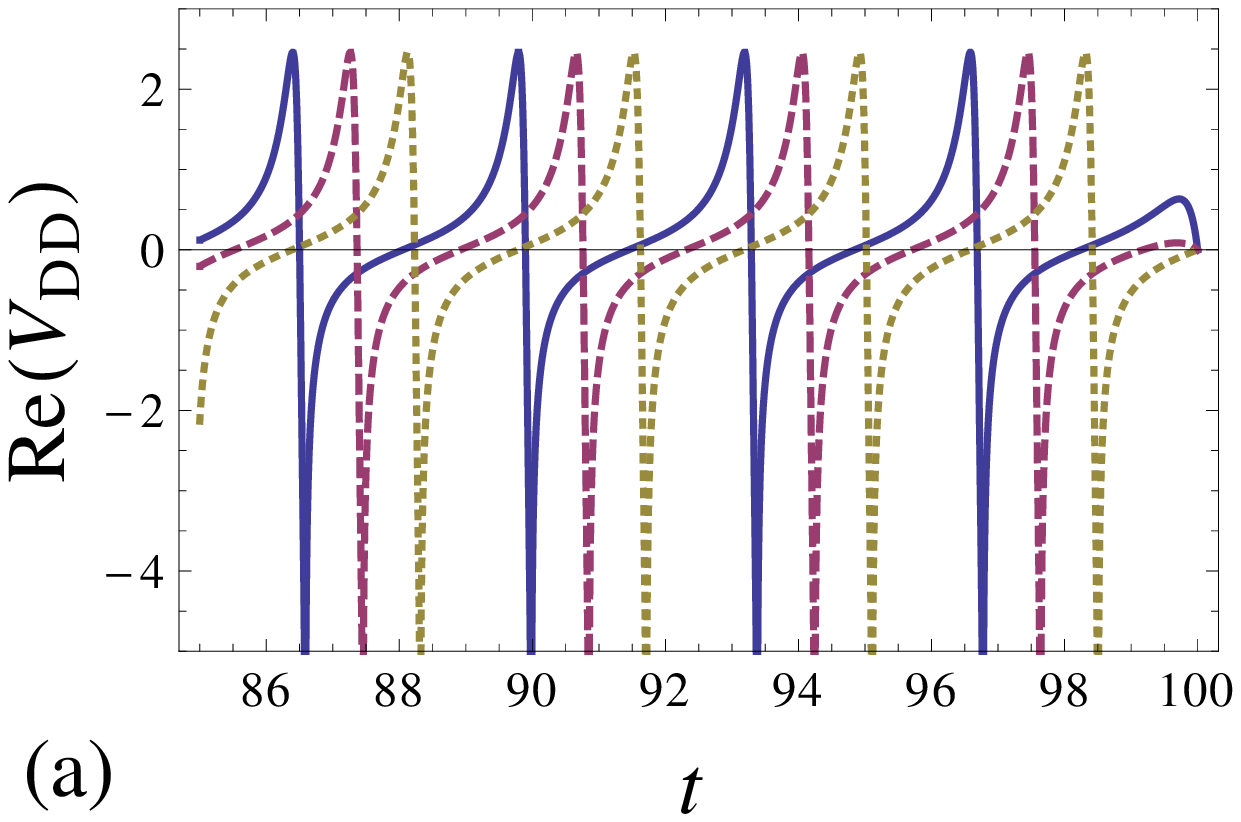}\includegraphics[clip,width=5.8cm]{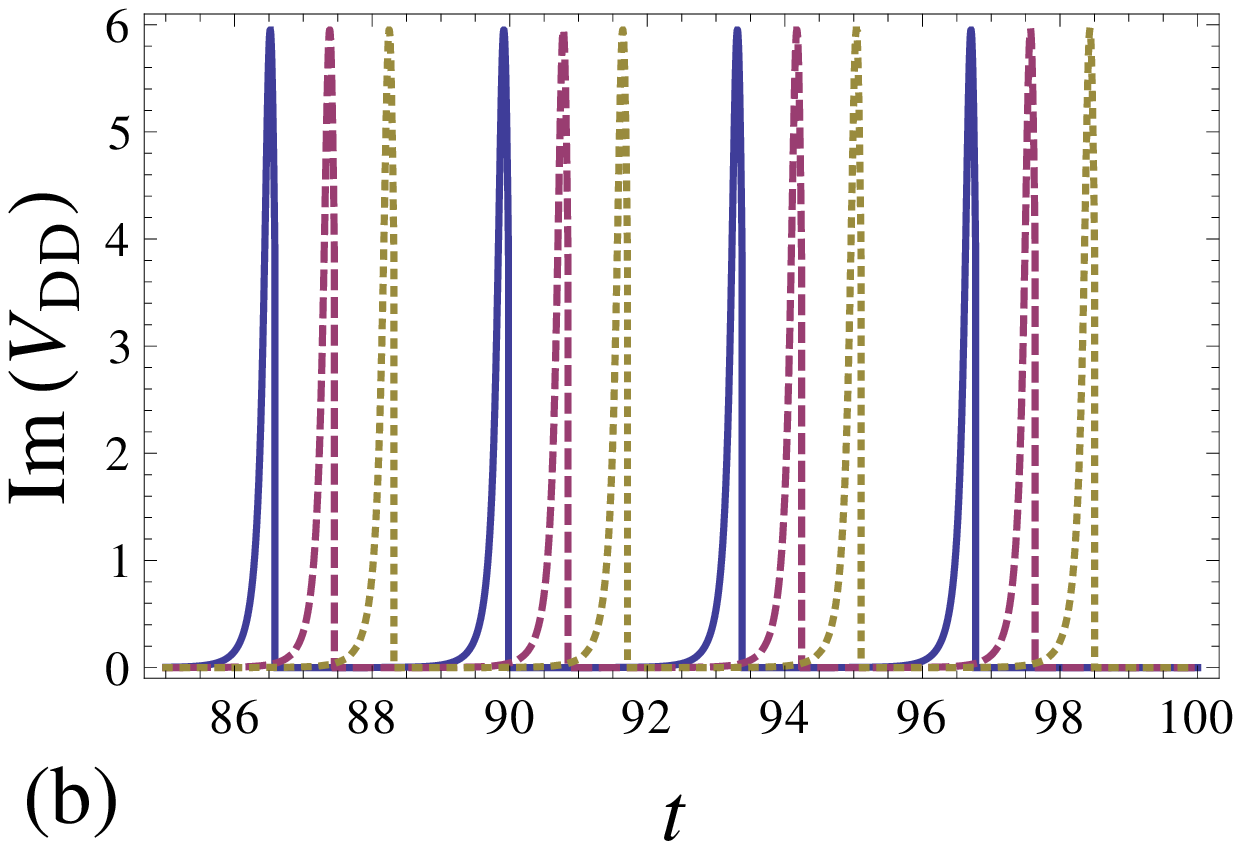}
\par\end{centering}
\caption{{(}Colour online{)} The evolution of $V_{\boson\boson}$ as a function of $t$ in the
scaling regime, starting with the initial condition $V_{\boson\boson}=0$
at $t=100$, for three different choices for the phase of $\Lambda^{\prime}$
on the asymptotic limit cycle (yellow dotted, red dashed, solid blue).
The imaginary part of $\Lambda^{\prime}$ at the starting point is
the same in the three cases shown. (a): Real part, (b): Imaginary part. \label{fig:VBB-scale}}
\end{figure}

\begin{figure}
\begin{centering}
\includegraphics[width=6cm]{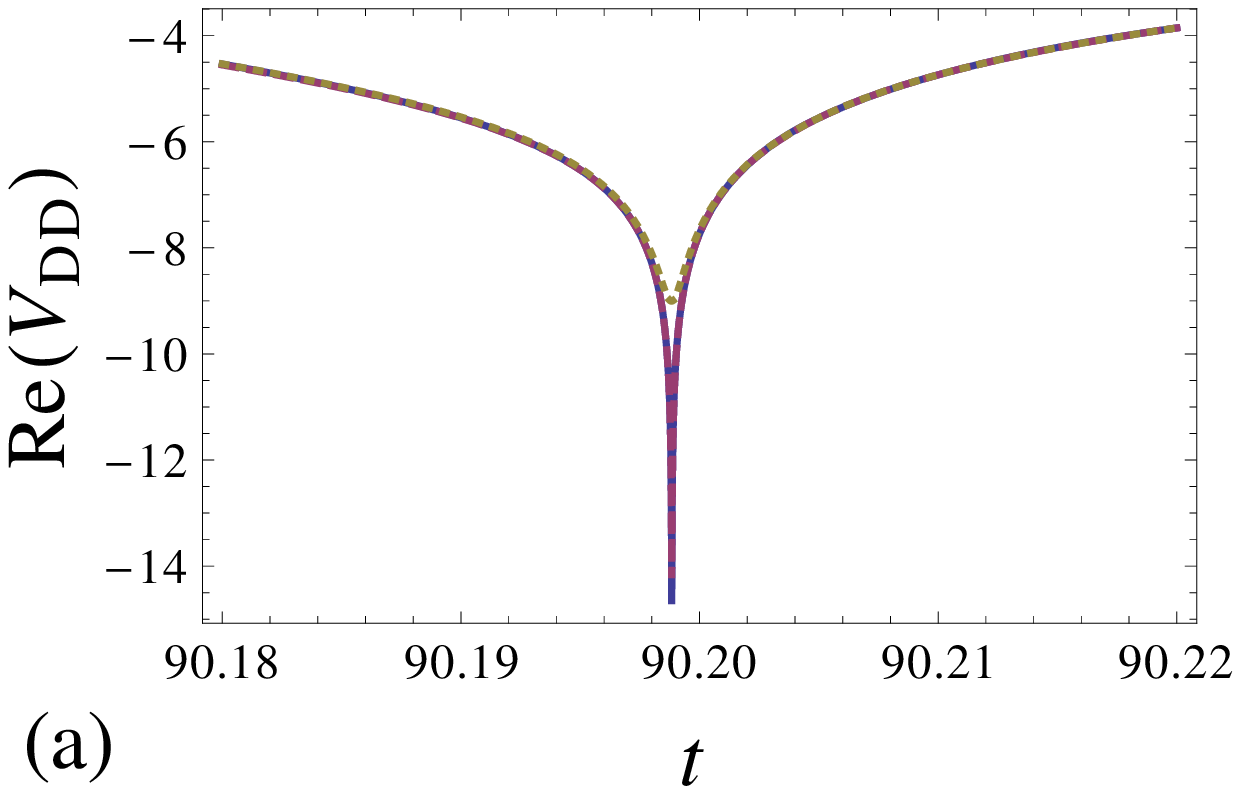}\includegraphics[clip,width=5.8cm]{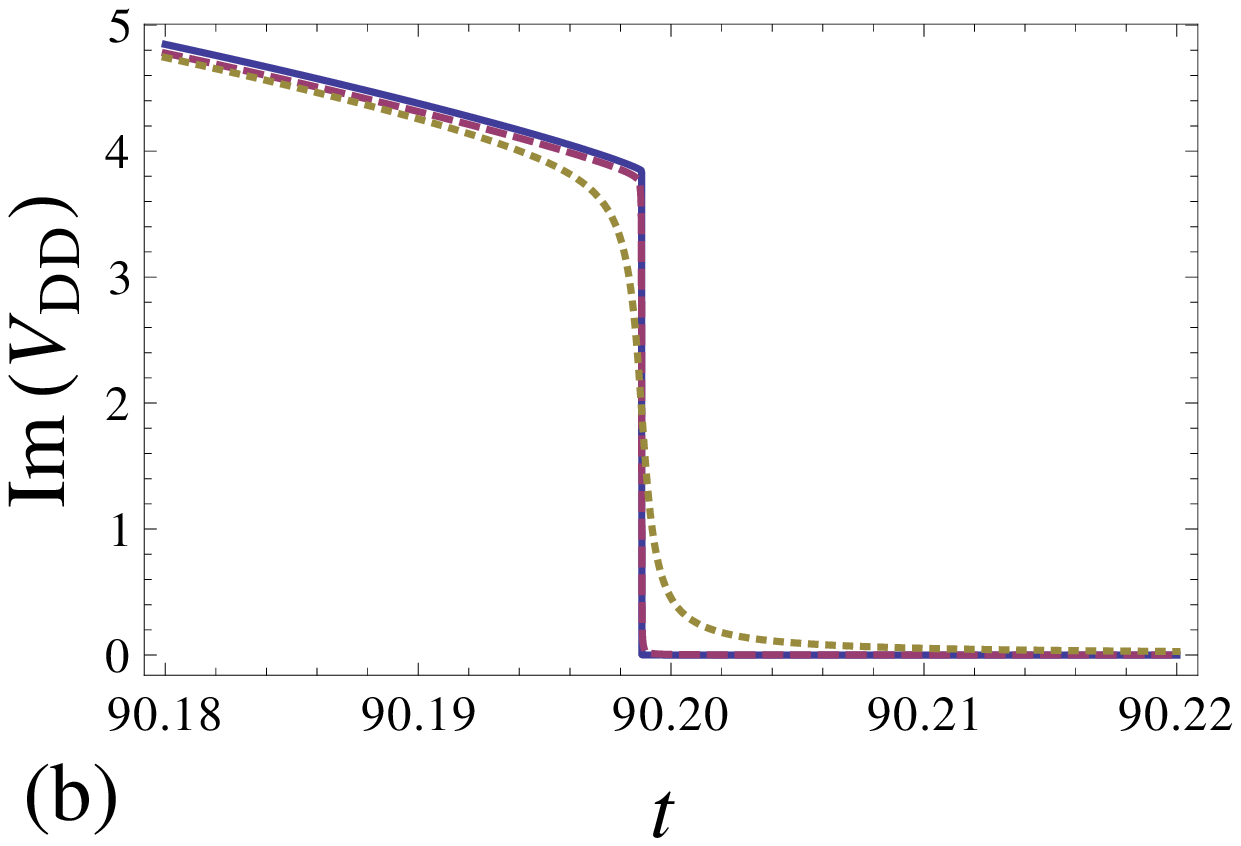}
\par\end{centering}
\caption{{(}Colour online{)} A single singularity of $V_{\boson\boson}$ in the scaling limit,
for three decreasing values $\delta=10^{-2}$, $10^{-4}$, $10^{-6}$
(yellow dotted, red dashed, solid blue)
of the imaginary part in the initial condition on $\Lambda^{\prime}$,
Eq.~(\ref{eq:initLam}). (a): Real part, (b): Imaginary part.
\label{fig:logsing}}
\end{figure}

The three-body parameter in spatially symmetric channel, $\lambda^{\prime}$, 
couples to the evolution of the dimer-dimer parameter in the corresponding
four-body channel, $\bar{u}_2$. The rescaled version of this, which we call 
$\bar{U}_{2}$, is defined as in Eq.~(\ref{eq:U2defn}). However in the region 
of large $\kappa \gg 1$, it is more convenient to work with 
\begin{equation}
V_{\boson\boson}(\kappa)=\kappa^{3}\bar{U}_{2}.
\end{equation}
The evolution of this in the scaling regime is shown in Figs.~\ref{fig:VBB-scale} 
and~\ref{fig:logsing}. There we can see that the solution, apart from a very 
short-lived transient, is completely determined by the driving term deriving from 
the coupling to $\Lambda^{\prime}$. The coupling $V_{\boson\boson}$ shows periodic
singularities, reflecting the Efimov physics in the three-body channel. The initial
transient effects vanish with a part of one Efimov cycle, reflecting the fact that the
exact boundary condition on $V_{DD}$ is an ``irrelevant" parameter.
More details of one of the singularities of $V_{DD}$ are shown in 
Fig.~\ref{fig:logsing}. As we decrease the imaginary part of
$\Lambda^{\prime}$, i.e., the value of $\delta$ in Eq.~(\ref{eq:initLam}),
we see that the solution behaves much like a logarithmic singularity,
with a sharp peak in its real part, and a finite jump in its imaginary part.

The rescaled coupling $V_{\boson\boson}$ goes to zero like $\kappa^{3}$
as $\kappa\rightarrow 0$. However we are ultimately interested in the
values of the unscaled couplings in this limit. This suggests
that the most appropriate technique for solving the differential 
equations numerically is to use the equations for the rescaled 
quantities, $V_{\boson\boson}$ and $\kappa^{2}\Lambda^{\prime}$ for 
$t=\ln\kappa>t_{0}\approx0$, and the equations for $\bar{U}_{2}$ and 
$\Lambda^{\prime}$ for $t<t_{0}$. Doing this we find that both couplings
tend to finite values in the physical limit. The physical value of
$\bar{U}_{2}$ is in general complex. This indicates that we are indeed 
looking at an inelastic channel, as a result of the more-deeply bound 
states in the three-body channel.

\begin{figure}
\begin{centering}
\includegraphics[clip,width=5cm]{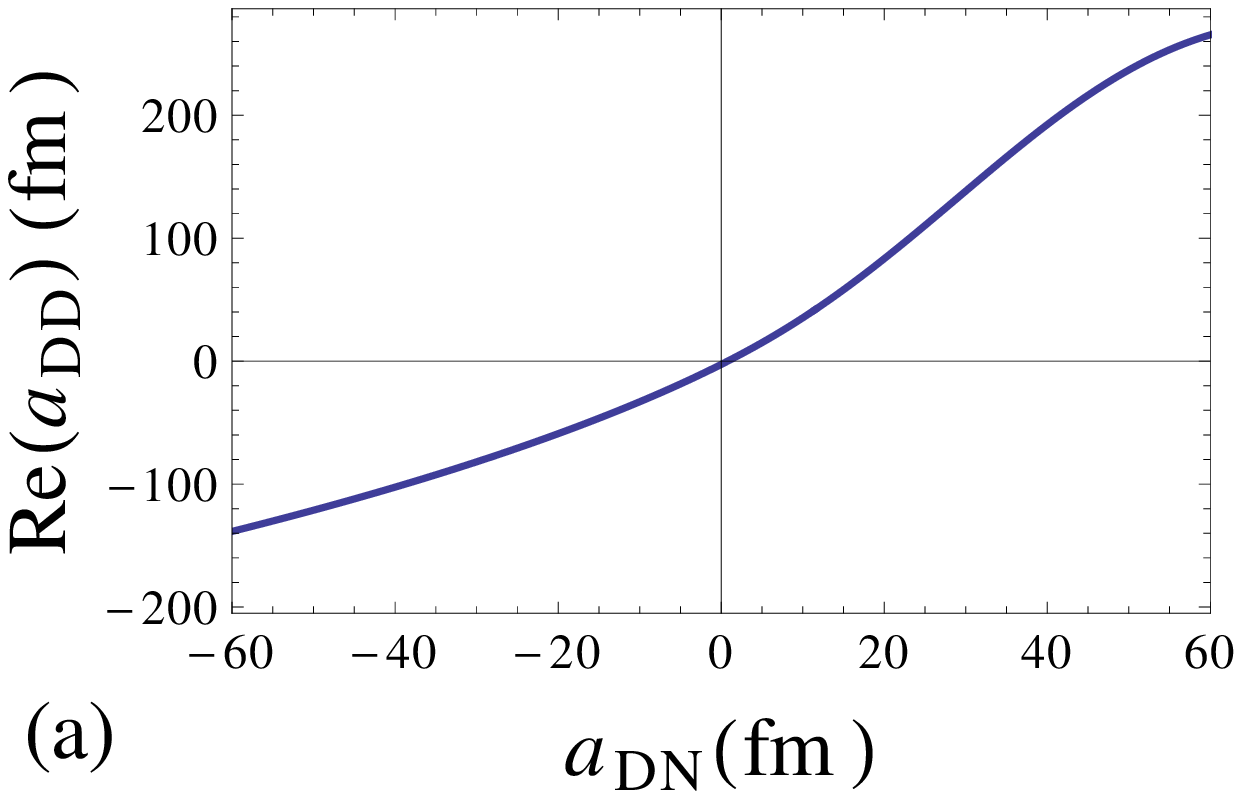}\includegraphics[clip,width=4.9cm]{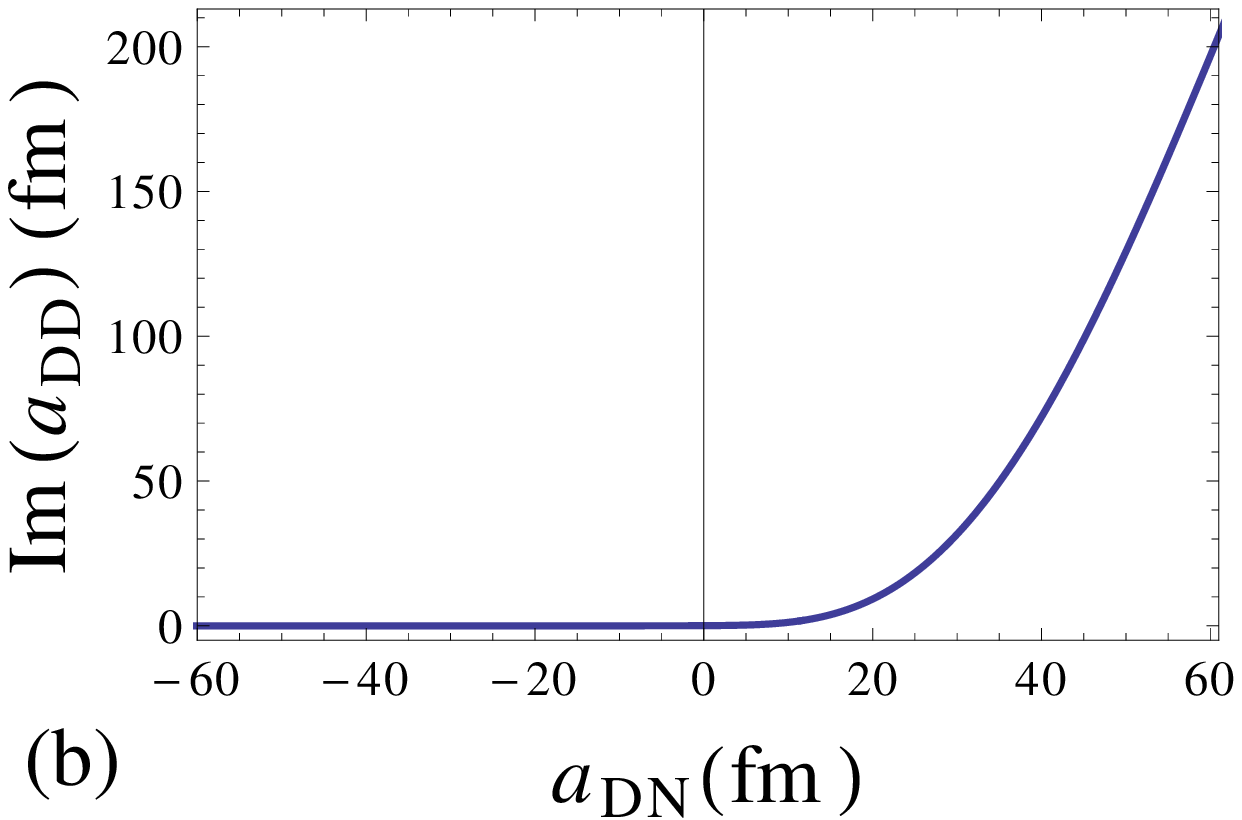}\includegraphics[clip,width=5cm]{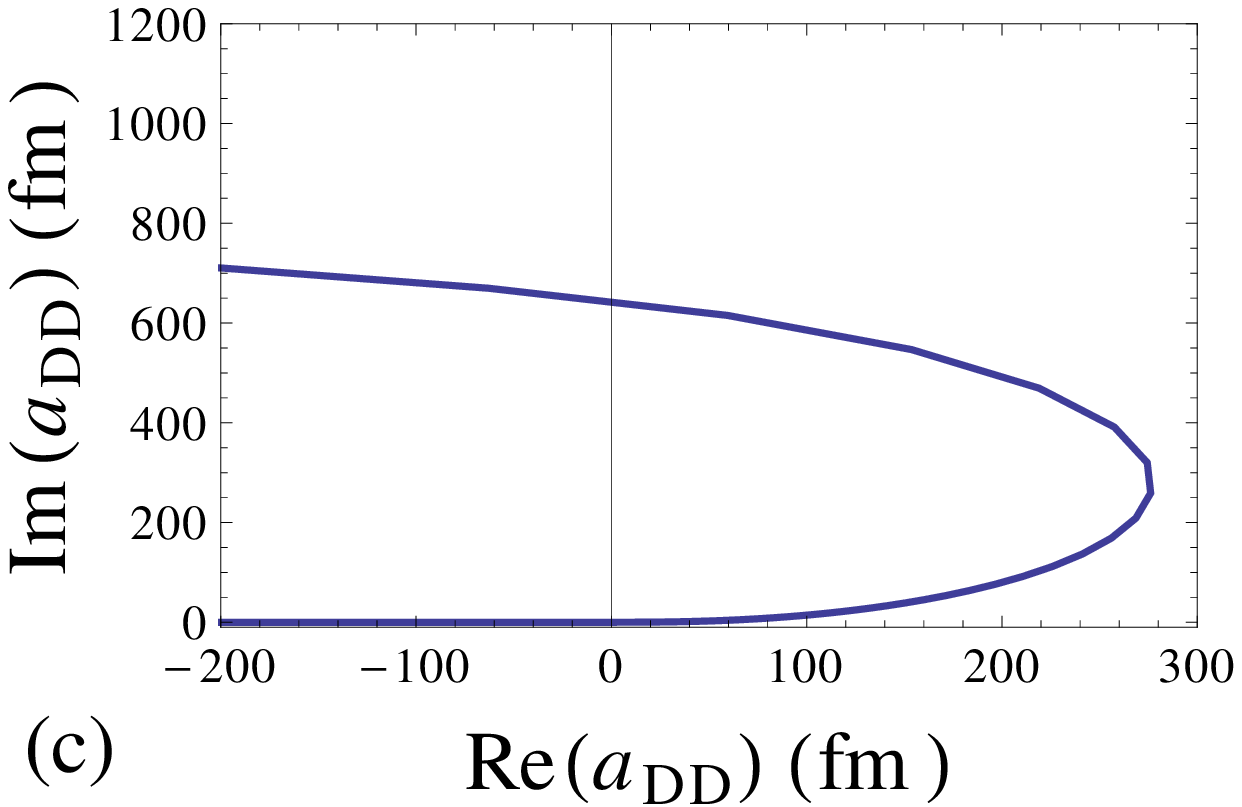}
\par\end{centering}
\caption{{(}Colour online{)} The relations between the scattering lengths in the SU(4) limit
for the channels exhibiting the Efimov effect. All the points on these
curves are obtained by choosing different points on the asymptotic
limit cycle for $\Lambda^{\prime}$ as initial values. 
(a): Real part of dimer-dimer scattering length; (b): Imaginary part of the same scattering length; (c): parametric plot of the real and imaginary parts.
{\label{fig:scatforcycle}}}
\end{figure}

Finally, we express our results in terms of scattering lengths, as discussed in 
Appendix B. We take the value $a_{\vb}=4.32\text{ fm}$ for the two-body system 
in order to reproduce the deuteron binding energy. This differs slightly from 
the experimental value because our current calculations do not include finite 
range corrections (i.e., momentum-dependent interactions). Using the relations  
\begin{eqnarray}
a_{\boson\boson}/a_{\vb} & = & 32\pi\bar{U}_{2}(0),\\
a_{\boson\fermion}/a_{\vb} & = & \frac{4}{3}\Lambda'(0),
\end{eqnarray}
we get the results summarised in Fig.~\ref{fig:scatforcycle}.  These
show the relationship between four-body scattering and the three-body
parameter associated with the limit-cycle of $\Lambda^{\prime}$, which
is fixed by the nucleon-dimer scattering length. There we see that the
real part of the dimer-dimer scattering length has a zero at
  almost the same energy as the the nucleon-dimer scattering
  length. We also see that there is a maximum value for the real part
of the $DD$ scattering length.

In the other pair of channels, which have spatial wave functions with mixed 
symmetry, all three- and four-body parameters are irrelevant. All the 
physical quantities can therefore be related to the single physical scale,
$a_\vb$ in our action. In particular, we find
\begin{eqnarray}
  a_{\boson\boson}^{\prime}/a_{\vb} & = & 32\pi U_{2}(0)=1.34,\\
  a_{\boson\fermion}^{\prime}/a_{\vb} & = & 
\frac{4}{3}\Lambda=\frac{5\sqrt{\frac{215}{7}}}{3}-\frac{29}{3}=-0.43.
\end{eqnarray}
The result that the dimer-nucleon scattering length in the
mixed-symmetry channel has the opposite sign to the the
nucleon-nucleon scattering length is a stable one. This is due to the
fact that we have a quadratic equation for the coefficient $\kappa^{-2}$
in the asymptotic behaviour of $\Lambda$, with a discriminant
$b^2-4ac$ where $a$, $b$ and $c$ are all positive--and we thus have
two negative solutions.  There are thus two asymptotic solutions, both
of which are negative.  (Details can be found in the discussion
surrounding Eqs.\ (\ref{eq:lam_as_large}) and (\ref{eq:lam_as_c0}).)
The sign of $\Lambda$ is preserved in the evolution out of the
asymptotic regime, and hence, even though the numerical details can
depend on the choice of cut-off function, as long as we insist on a
consistent $k$-scaling of the cut-off function the opposite sign
remains.

\subsection{Broken SU(4)}

In the real world, the singlet and triplet nucleon-nucleon scattering
lengths are different and there is no exact SU(4) symmetry. The
evolution equations do not decouple and so we have deal with $2\times
2$ matrices of coupling constants in both the $(1/2,1/2)$
three-nucleon channels and the $(0,0)$ four-nucleon ones. In each case
we can identify two scattering ``eigenchannels'' and, from the
zero-energy $T$ matrix, determine a scattering length in each of these
channels. We show the general behaviour of these scattering lengths
here, taking 
the value of $1-\alpha=1-a_{\vb}/a_{\ib}=5/4$
for the SU(4)
breaking parameter. This is close to the realistic case, giving
$a_{\ib}=-17.2\text{ fm}$. 
For example, for the $A=3$
  $(1/2,1/2)$ channels we must diagonalise the matrix of couplings (see Eqs.\
  (\ref{eq:lamvv},\ref{eq:lamvi}))
\begin{equation}
\begin{pmatrix}
\lambdahh_{\vsum\vsum}&\lambdahh_{\vsum\isum}\\
\lambdahh_{\isum\vsum}&\lambdahh_{\isum\isum}
\end{pmatrix}.
\end{equation}
For exact SU(4) symmetry, these couplings are all equal and the
eigenvectors are just $(1,1)$ and $(1,-1)$. More generally,
transforming the matrix into this basis gives the SU(4)-symmetric and
SU(4)-breaking couplings shown in
Eqs. (\ref{eq:lamSU4}--\ref{eq:deltalamSU4}). In a similar way, the
SU(4) eigenchannels for the $A=4$ $(0,0)$ case are also equal
mixtures.

\begin{figure}
\begin{centering}
\includegraphics[width=8cm]{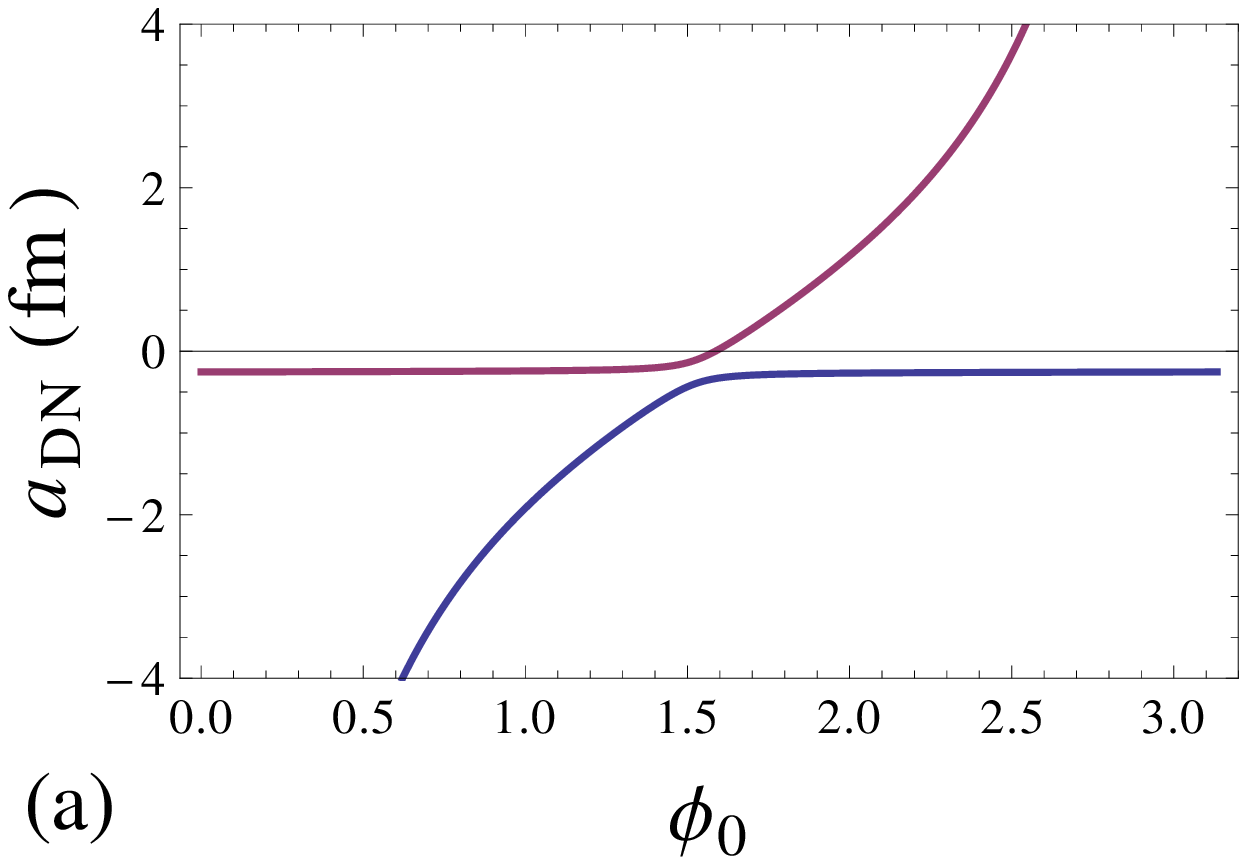}\\
\includegraphics[width=8cm]{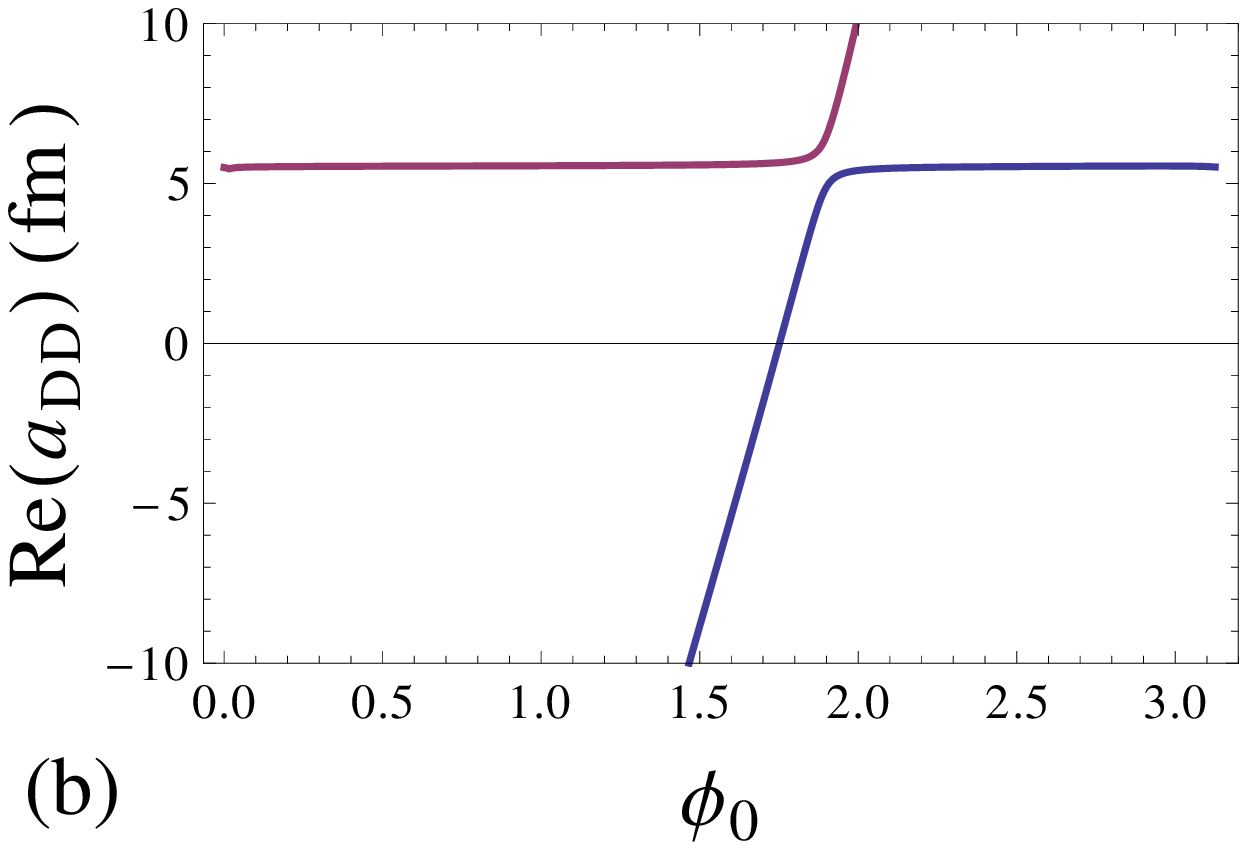}\includegraphics[width=8cm]{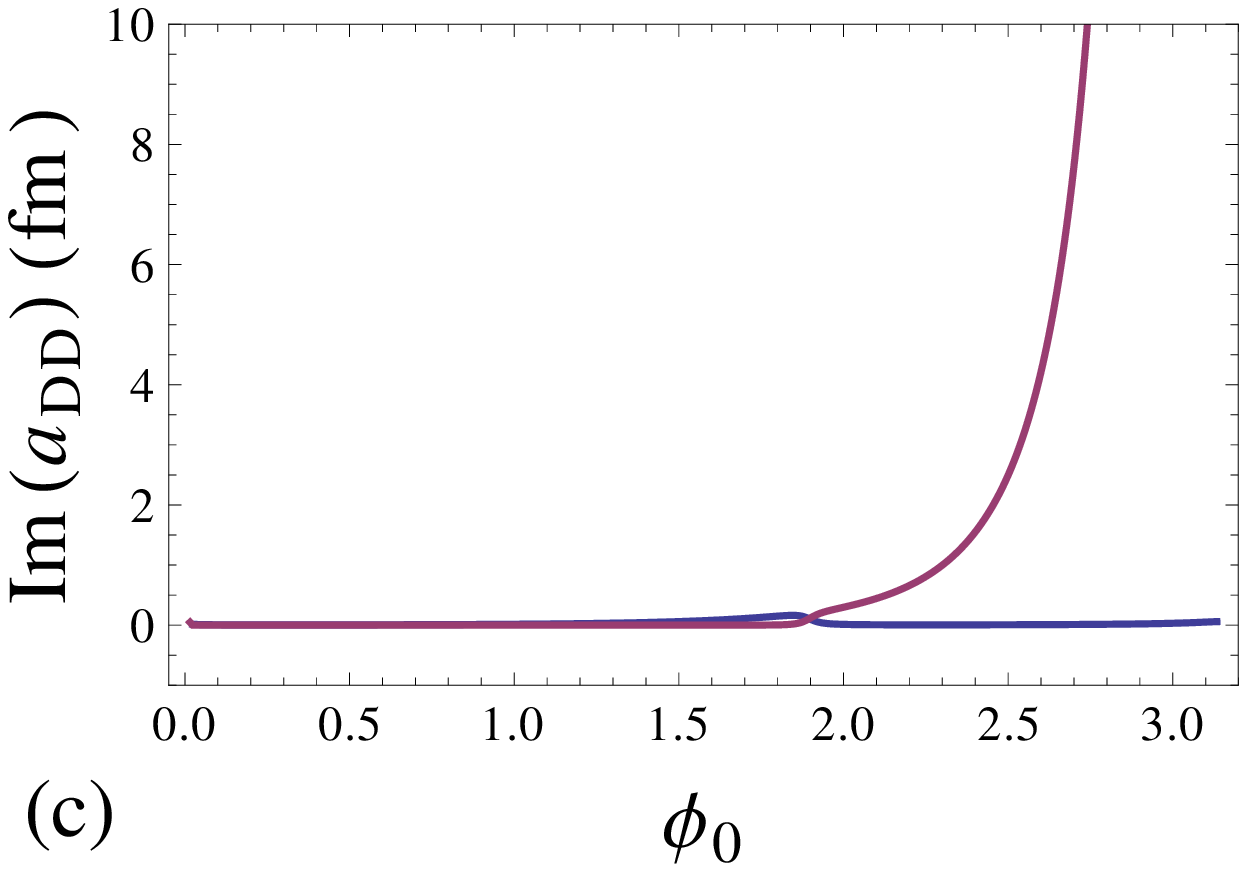}\\
\includegraphics[width=8cm]{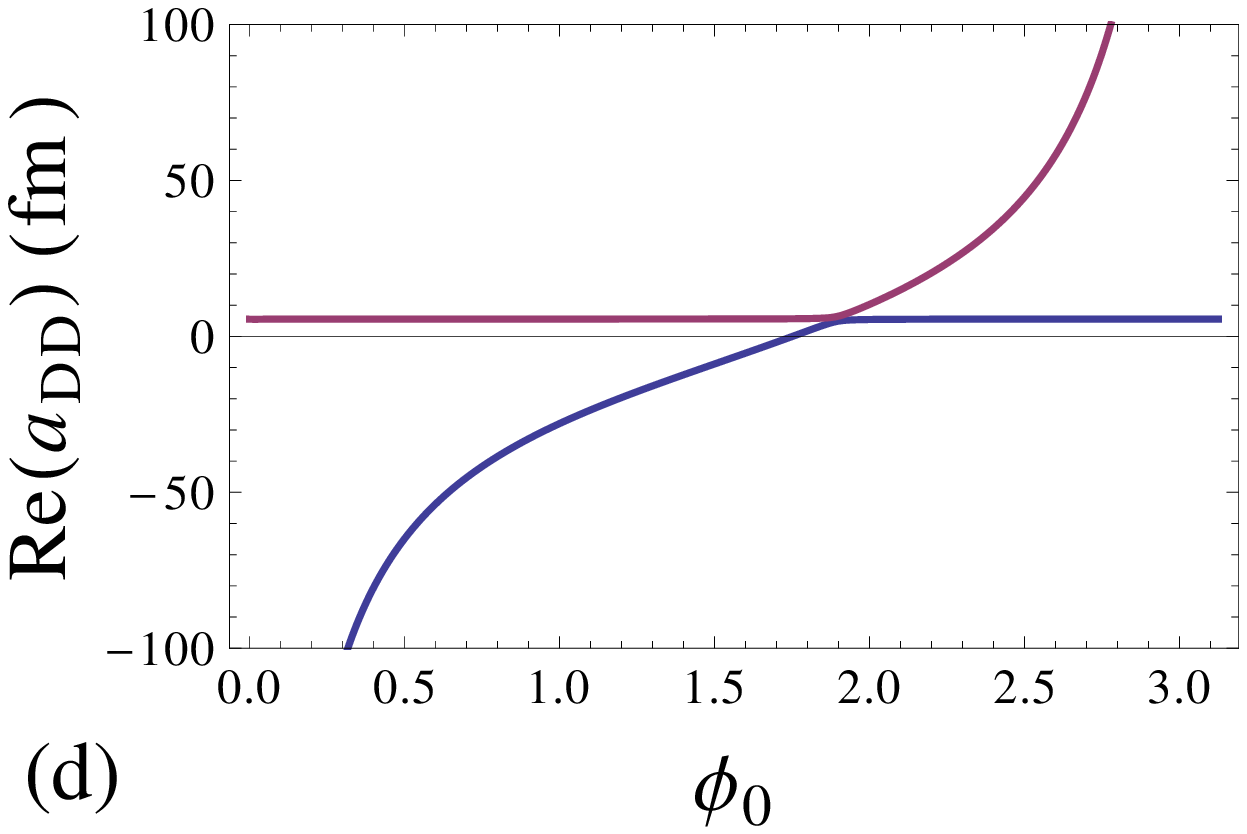}
\par\end{centering}
\caption{{(}Colour online{)} Evolution of the eigenvalues of the $T$-matrix (parametrised as scattering
lengths) in the two $A=3$ $(1/2,1/2)$ eigenchannels, and similar for
the two $A=4$ $(0,0)$ eigenchannels, as a function of the parameter $\phi_{0}$
specifying the initial condition on the limit cycle. 
(a): Nucleon-dimer scattering length (which is real), (b): Real part of the dimer-dimer scattering lengths (fine scale), (c) Imaginary part of the dimer-dimer scattering lengths, (d): Real part of the same on a course scale, showing the narrowness of the avoided crossing.
\label{fig:scatl2}}
\end{figure}

\begin{figure}
\begin{centering}
\includegraphics[clip,width=8cm]{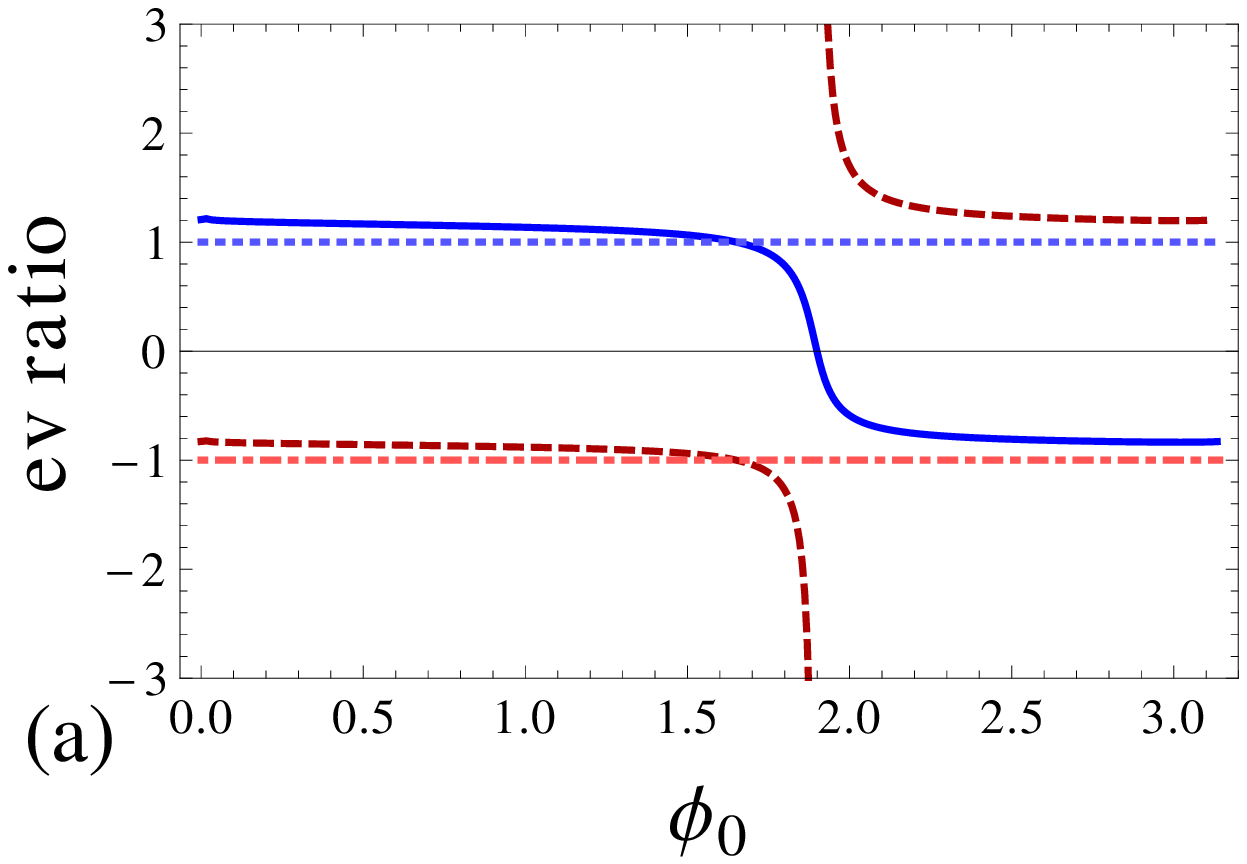}\includegraphics[width=8cm]{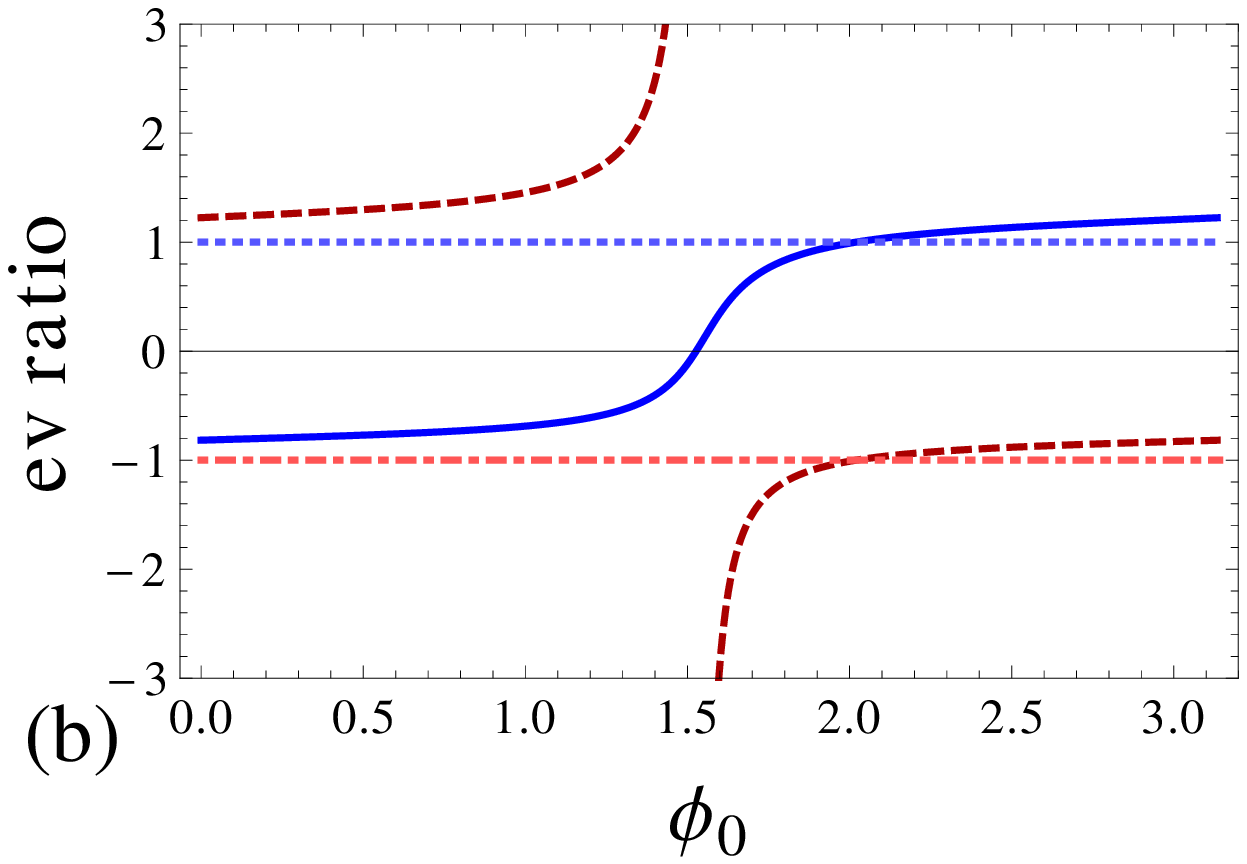}
\par\end{centering}
\centering{}\caption{{(}Colour online{)} Ratio between components of
  each of the two eigenchannel solutions found in the cases
    with coupled channels. As discussed in the text, if this ratio is
    close to $\pm 1$ we have the perfect mixing expected in the SU(4)
    symmetric limit.  (a): $\lambdahh$ and (b):
  $\bar{u}_{2}$.\label{fig:evoratio} }
\end{figure}

The results depend on the starting on the limit cycle of the coupling constant 
that displays the Efimov effect in the scaling limit. The dependencies of the
scattering lengths on this three-body parameter are presented in 
Fig.~\ref{fig:scatl2}. Both lengths show avoided crossings; the narrowness 
of these is an indication that SU(4)-breaking effects are relatively weak.
Further evidence of the smallness of this breaking is provided by the ratios between the components of the 
eigenchannel solutions. As can been seen from Fig.~\ref{fig:evoratio}, the
mixings are small ($\lesssim 20$\%), except in narrow windows around
the crossing points.

\begin{figure}
\begin{centering}
\includegraphics[width=8cm]{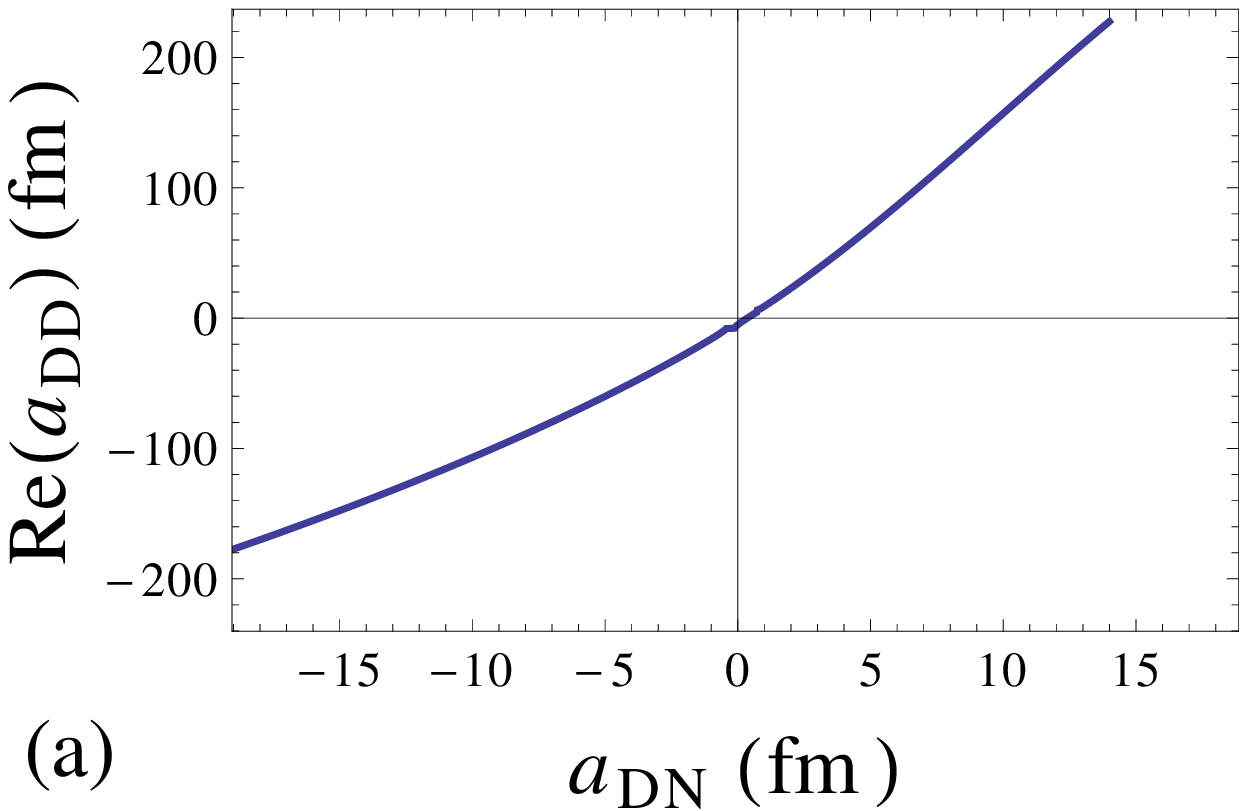}\includegraphics[width=8cm]{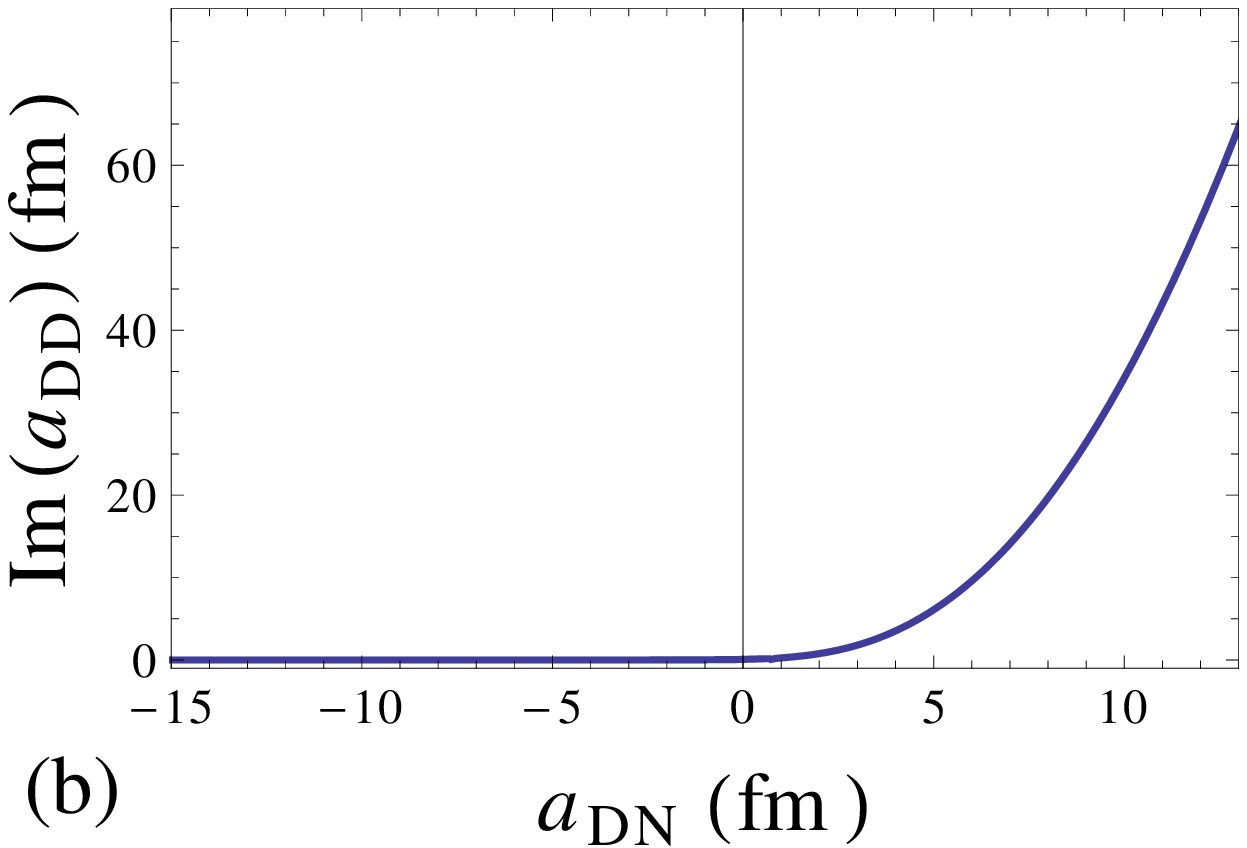}
\par\end{centering}
\begin{centering}
\includegraphics[width=8cm]{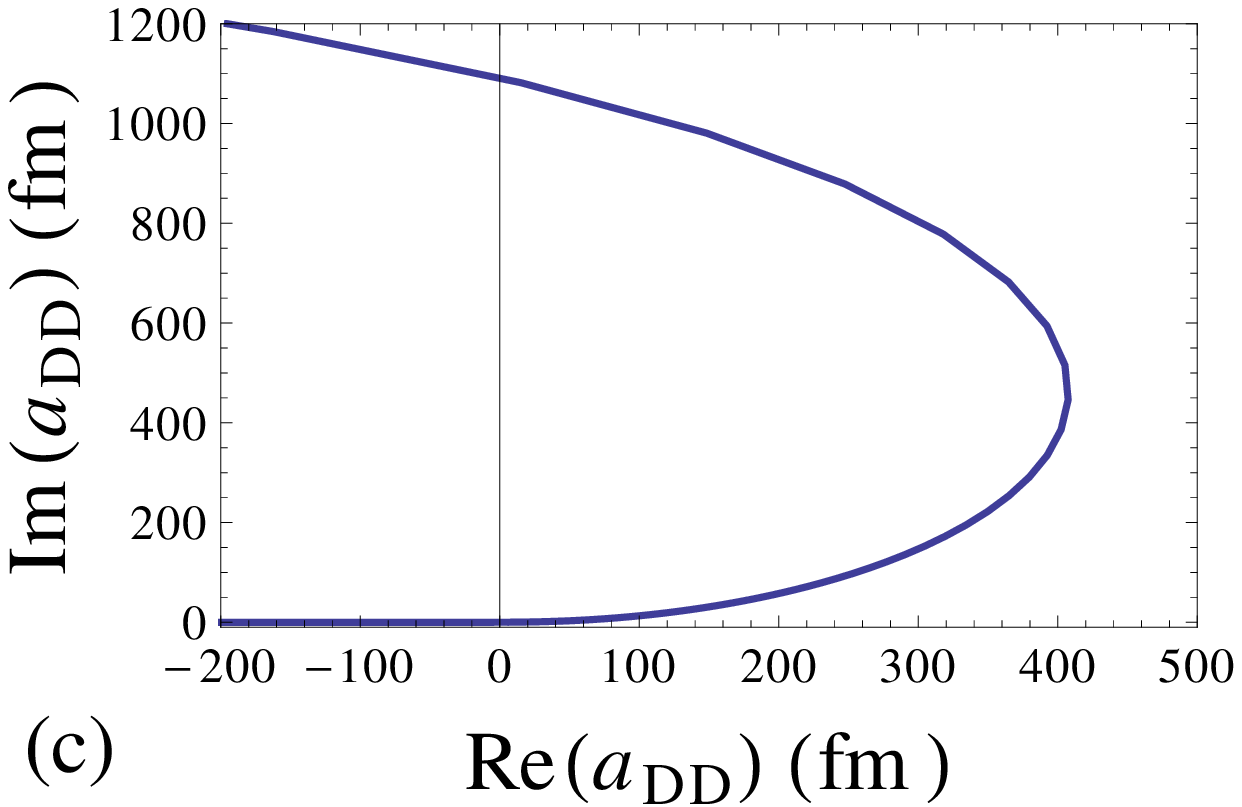}
\par\end{centering}
\caption{{(}Colour online{)} The relations between the scattering lengths in the Efimov-dominated 
channels for $\alpha=-1/4$.
(a): Real part of dimer-dimer scattering length; (b): Imaginary part of the same scattering length; (c): parametric plot of the real and imaginary parts.
{\label{fig:scatforcycle-1}}}
\end{figure}

In Fig.~\ref{fig:scatforcycle-1} we show the relations between the scattering lengths
in the channels dominated by the Efimov effect. The scattering length in these channels  vary rapidly with the three-body
parameter. This means that at the avoided crossing, we switch from one branch of the solution 
to the other. As a result there is a very small discontinuity in the plots, close to the points 
where $a_{\boson\fermion}$ and $\Re[a_{\boson\boson}]$ vanish.
In the other channels we have an approximately constant scattering lengths away from the avoided
crossing. These have values of $a_{\boson\fermion}=-0.26\pm0.03\:\text{fm}$ and 
$\Re[a_{\boson\boson}]=5.55\pm0.11\ \text{fm}$.

\subsection{Realistic scattering lengths}

We now turn to the implications of these results  for realistic
few-nucleon systems. Above we have already shown some results obtained with 
values for the nucleon-nucleon scattering lengths that are close to the 
observed ones. The small differences are not significant given the level
of truncation of our REA, which means that effective-range corrections are 
not included. 

As already discussed, we also need one piece of three-body data to fix
the starting point of the evolution on the Efimov cycle that is
present for large cut-off scales. We choose to do this by using the
experimental value of the spin-doublet $nd$ scattering length
$^2a_{nd} = 0.68$~fm \cite{Dil} to determine the initial value of
$\lambda_{\vb\vb}^{(1/2,/1/2)}$ (since $nd$ is not one of the
scattering eigenchannels). We have to be careful here, since, as
stated earlier, we calculate the three-body couplings off-shell, at
energy $\ed/2$ below threshold.

The situation is different for the $nd$ scattering length $^4a_{nd}$
in the spin-quartet channel. All the three-body parameters that
contribute here are irrelevant and so the physical value of $^4a_{nd}$
extracted at $k=0$ does not depend on their initial values. Our FRG
calculations give $^4a_{nd}=-1.02\ \text{fm}$. This differs
significantly from the experimental value of $^4a_{nd}=6.35\
\text{fm}$ \cite{Dil}.  Other theoretical calculations seem to be able
to reproduce this result, either by applying the $\epsilon$ expansion \cite{Rup} or by solving the
Skornyakov--Ter-Martirosyan equation \cite{Birra}. As we have said
before, the negative value of this scattering length appears to be a
stable result of our calculations, and is not very sensitive to the
fine details. One possible reason for this is that the value of
$^4a_{nd}$ reported here is defined not at threshold but at the
expansion point we have used for the nucleon energy in our flow
equations, $\mathcal{E}_{D}/2\approx-1\ \text{ MeV}$.  Energy
dependence is known to be important for three-body observables
\cite{Die} and so that the extrapolation to the physical threshold
could have a significant effect, especially since $\ed$ is a small
scale, we could expect significant energy dependence. The work of
Bedaque \emph{et al} \cite{Birra} shows energy dependence, but when
they include only the scattering length, this would probably not be
sufficient to explain what happens here.  In order to explore
this in more detail, we will need to extend our REA to include
energy/momentum-dependent couplings.

In the case of four-nucleon systems, we consider two interacting deuterons 
and extract the $dd$ scattering lengths in the spin-singlet and quintet 
channels. Besides being of interest on its own, the low-energy $dd$ 
interaction may have astrophysical applications, since under conditions 
expected to exist inside brown-dwarf stars a many-deuteron system may behave 
as a superfluid \cite{Bed}. A framework like the FRG that can describe
both the $dd$ interaction in vacuum and be extended to dense matter
would be very useful in this context.

In the case of two deuterons, our expansion point corresponds to the 
physical threshold and so issues of extrapolation in energy do not arise.
The singlet channel can couple to the $n+^3$He (or $p+^3$H) channel
which has a lower threshold. Hence, as we have already seen above, the 
scattering can be inelastic, and its scattering length is complex.
In contrast, the coupling of the quintet channel to the rearrangement
ones is much smaller as non-zero orbital angular momentum is required
\cite{Fil}. In our $S$-wave treatment, this channel is closed.

Our result for $\bar{u}_{2}$ corresponds to a singlet $dd$ scattering
length with $\Re[^1a_{dd}]=4.44\ \text{fm}$ and 
$\Im[^1a_{dd}]=0.17\ \text{fm}$. The real part is consistent 
with the value of $\Re[^1a_{dd}]=4.9\ \text{fm}$ obtained in 
Ref.~\cite{Car} by solving the Faddeev--Yakubovsky equation, even though the 
imaginary part is larger. 

In the spin-quintet $dd$ channel we get $\Re[^5a_{dd}]=2.55\
\text{fm}$.  This also agrees with the value $\Re[^5a_{dd}]=3.2\
\text{fm}$ obtained by Rupak in the framework of an $\epsilon$
expansion \cite{Rup}. This is perhaps unsurprising since that work
uses an effective field theory with a Lagrangian that like our REA
omits effective-range terms. What is more puzzling is that Rupak finds
a very different value from ours for $^4a_{nd}=4.78\ \text{fm}$, the
scattering length in the three-body channel that feeds into the
$(2,0)$ four-body one. This may be yet another suggestion that energy
dependence of these couplings, allowing us to extrapolate better to the on-shell value, needs to be considered.

Our result for the deuteron-deuteron scattering length in the quintet
channel agree qualitatively with the exact quantum mechanical analysis
in Ref.~\cite{Fil}, given our incomplete treatment of the 3+1-particle
rearrangement channels. The authors of that work find that this
scattering length is very sensitive to these channels, and excluding them
leads to a substantial reduction of $\Re[^5a_{dd}]$, from $7.5\ \text{fm}$ to
$-0.1\ \text{fm}$.

In this first application of the FRG method to three- and four-nucleon 
systems, we are able to reproduce some of the effects of inelasticities 
in deuteron-deuteron scattering. However at this level we do not 
describe the nucleon-trimer threshold that is needed to get the
correct energy dependence in this channel. To do this we would 
need to extend our approach by adding an auxiliary trimer field,
along the lines suggested by Schmidt and Moroz \cite{Sch}.
This is certainly feasible but it would require adding a number 
of additional interaction terms to our REA, leading to a
much larger set of coupled evolution equations.

\section{Conclusions}

In summary, we have applied the FRG method to three- and four-nucleon
systems, both in the limit of exact Wigner SU(4) symmetry, and with
realistic symmetry breaking. From the couplings in the physical limit,
we have calculated the nucleon-deuteron and deuteron-deuteron
scattering lengths in various spin-isospin channels.

We find that the evolution of one three-body coupling shows
oscillatory, limit-cycle behaviour which is a manifestation of the
Efimov effect in the corresponding channel. We therefore need to use
one piece of data to fix one three-body parameter; all other three-
and four-nucleon observables in the spatially symmetric are then
predicted in terms of this and the two-body scattering lengths. In
contrast, the observables in the channels with mixed spatial symmetry
are independent of the initial scale, provided it is chosen large
enough. Their values are therefore determined by the two-body input
alone.

We have explored the dependence of the three- and four-body
  couplings on the three-body parameter. We find that the pattern of
  physical couplings remains within about 20\% of the SU(4) limit,
  except in rather narrow windows of the parameter. This is reflects
  the fact that the Efimov effect introduces an additional momentum
  scale in few-nucleon systems, associated with the lowest-energy
  three-body bound states. Provided that these states are more deeply
  bound than the deuteron, as they are in the real world, this scale
  is large compared to the inverse scattering lengths, and SU(4)
  remains a good approximate symmetry.

The Efimov effect appears in the spin-doublet nucleon-deuteron and the
singlet deuteron-deuteron channels. Using the doublet scattering
length to fix the three-body parameter, we get a value for the singlet
deuteron-deuteron scattering length that is very close to one obtained
in the exact quantum mechanical calculations. In contrast, our value
for the spin-quintet scattering length shows no significant dependence
on the three-body parameter, but differs from the results of other
calculations, potentially as a result of ignoring off-shell effects in
this work.

There is a variety of ways in which the present study could be
improved.  One important one is the introduction of a trimer field in
order to describe better the nucleon-trimer channel, which is
responsible for the inelasticity in deuteron-deuteron scattering.
Even though this would remove some of the restrictions faced
  in this work, it is equally important to extend the ansatz for the
  running action to include energy- or momentum-dependent couplings.
  The first would allow the calculation of scattering observables away
  from (unphysical) thresholds, which would allow us to better compare
  scattering lengths to physical results.  The second would allow
  inclusion of effective-range corrections.  Investigations of all of
  these extensions are under way.

\section*{Acknowledgements}

The work of one of the authors (BK) was supported by the EU FP7
programme (Grant 219533). MCB acknowledges useful discussions with
S. Moroz and R. Schmidt.

\appendix

\section{SU(4) limit}

In the limit of exact SU(4) symmetry, $a_{\vb}=a_{\ib}$, $\alpha=1$,
the problem simplifies considerably. None of the loop integrals now
depends on the type of dimer considered. Looking at the evolution
equations it becomes quickly obvious that it is useful to introduce
new constants,\begin{eqnarray}
\frac{1}{2}\left(\lambda+\lambda^{\prime}\right) & = & \lambdahh_{\vsum\vsum}=\lambdahh_{\isum\isum},\\
\frac{1}{2}\left(\lambda-\lambda^{\prime}\right) & = & \lambdahh_{\vsum\isum}=\lambdahh_{\isum\vsum},\\
\lambda^{\prime\prime} & = & \lambdath_{\vsum\vsum}=\lambdaht_{\isum\isum},\\
\bar{u}_{2}^{\prime} & = & \bar{u}_{2\vsum\vsum}+\bar{u}_{2\vsum\isum},\\
\bar{u}_{2} & = & \bar{u}_{2\vsum\vsum}-\bar{u}_{2\vsum\isum},\end{eqnarray}
which satisfy much simpler equations than the {}``channel couplings''.
The $ND$ coupling constants satisfy\begin{eqnarray}
\partial_{k}\lambda & = & I_{1}\lambda^{2}+g^{2}I_{2}\lambda+\frac{1}{4}g^{4}I_{3},\label{eq:lamSU4_2}\\
\partial_{k}\mbox{\ensuremath{\lambda}}^{\prime} & = & I_{1}\lambda^{\prime2}-2g^{2}I_{2}\lambda^{\prime}+g^{4}I_{3},\label{eq:lampSU4_2}\\
\partial_{k}\lambda^{\prime\prime} & = & I_{1}\lambda^{\prime\prime2}+g^{2}I_{2}\lambda^{\prime\prime}+\frac{1}{4}g^{4}I_{3}.\label{eq:lamppSU4}\end{eqnarray}
Since the first and last equation are identical, we thus realise that
$\lambda=\lambda^{\prime\prime}$. The potential then simplifies to
\begin{eqnarray}
\mathcal{V}_{3,BF} & = & \lambda^{\prime}\left[\vb^{\dagger}(p_{30},\vec{p}_{3})\psi^{\dagger}(p_{10},\vec{p}_{1})-\ib^{\dagger}(p_{30},\vec{p}_{3})\psi^{\dagger}(p_{10},\vec{p}_{1})\right]^{(1/2,1/2)}\cdot\nonumber \\
 &  & \quad\left[\psi(p_{20},\vec{p}_{2})\vb(p_{40},\vec{p}_{4})-\psi(p_{20},\vec{p}_{2})\ib(p_{40},\vec{p}_{4})\right]^{(1/2,1/2)}\nonumber \\
 &  & +\lambda\left[\vb^{\dagger}(p_{30},\vec{p}_{3})\psi^{\dagger}(p_{10},\vec{p}_{1})+\ib^{\dagger}(p_{30},\vec{p}_{3})\psi^{\dagger}(p_{10},\vec{p}_{1})\right]^{(1/2,1/2)}\cdot\nonumber \\
 &  & \quad\left[\psi(p_{20},\vec{p}_{2})\vb(p_{40},\vec{p}_{4})+\psi(p_{20},\vec{p}_{2})\ib(p_{40},\vec{p}_{4})\right]^{(1/2,1/2)}\nonumber \\
 &  & +\lambda\left[\ib^{\dagger}(p_{30},\vec{p}_{3})\psi^{\dagger}(p_{10},\vec{p}_{1})\right]^{(3/2,1/2)}\cdot\left[\psi(p_{20},\vec{p}_{2})\ib(p_{40},\vec{p}_{4})\right]^{(3/2,1/2)}.\end{eqnarray}
The $u_{2}$ equations also simplify\begin{eqnarray}
\partial_{k}u_{2} & = & \frac{1}{2}u_{2}^{2}K_{1}-2g^{2}\lambda^{\prime\prime}K_{2}-\frac{3}{4}g^{4}K_{3},\label{eq:u2SU4}\\
\partial_{k}u_{2,\vsum\isum} & = & \frac{1}{2}u_{2,\vsum\isum}^{2}K_{1}-2g^{2}\lambda K_{2}-\frac{3}{4}g^{4}K_{3},\label{eq:u2vtSU4}\\
\partial_{k}\bar{u}_{2}^{\prime} & = & \frac{1}{2}\bar{u}_{2}^{\prime2}K_{1}-2g^{2}\lambda K_{2}-\frac{3}{4}g^{4}K_{3},\label{eq:u2bpSU4}\\
\partial_{k}\bar{u}_{2} & = & \frac{1}{2}\bar{u}_{2}^{2}K_{1}-2g^{2}\lambda^{\prime}K_{2}+\frac{3}{2}g^{4}K_{3}.\label{eq:u2bSU4}\end{eqnarray}
Thus we also conclude that $u_{2,\vsum\isum}=u_{2}=\bar{u}_{2}^{\prime}$.
The potential terms in this limit can thus be written as\begin{eqnarray}
\mathcal{V}_{2,BB} & = & \frac{1}{2}u_{2}\biggl(\vec{\vsum}^{\dagger}(p_{10},\vec{p}_{1})\cdot\vec{\vsum}(p_{30},\vec{p}_{3})+\vec{\isum}^{\dagger}(p_{10},\vec{p}_{1})\cdot\vec{\isum}(p_{30},\vec{p}_{3})\biggr)\nonumber \\
 &  & \qquad\biggl(\vec{\vsum}^{\dagger}(p_{20},\vec{p}_{2})\cdot\vec{\vsum}(p_{40},\vec{p}_{4})+\vec{\isum}^{\dagger}(p_{20},\vec{p}_{2})\cdot\vec{\isum}(p_{40},\vec{p}_{4})\biggr)\nonumber \\
 &  & +\frac{1}{12}(\bar{u}{}_{2}-u_{2})\left[\vec{\vb}^{\dagger}(p_{10},\vec{p}_{1})\cdot\vec{\vb}^{\dagger}(p_{20},\vec{p}_{2})-\vec{\ib}^{\dagger}(p_{10},\vec{p}_{1})\cdot\vec{\ib}^{\dagger}(p_{20},\vec{p}_{2})\right]\nonumber \\
 &  & \qquad\left[\vec{\vb}(p_{30},\vec{p}_{3})\cdot\vec{\vb}(p_{40},\vec{p}_{4})-\vec{\ib}(p_{30},\vec{p}_{3})\cdot\vec{\ib}(p_{40},\vec{p}_{4})\right],\end{eqnarray}
a simple sum of the non-local generalisations of the square of the
linear and the quadratic Casimir invariant of $U(4)$.

\subsection{Unitary limit}

It is  instructive to look at what happens for large $\kappa=ka_{0}$.
This corresponds to the unitary limit, where the scattering length
diverges. The behaviour for large $\kappa$ is universal, and thus
provides boundary conditions for the numerical solution of the full
problem.

\subsubsection{Three-body coupling}

In the three-body case we are left with a set of dimensionless integrals (the tilde denotes we have scaled the expressions give before by appropriate powers of 
$M$ and $a_{\vb}$)
\begin{eqnarray}
\tilde{I}_{1} & = & \frac{16\kappa^{2}\left(\kappa^{2}+1\right)\left(45\left(\kappa^{2}+1\right)^{3}\cot^{-1}(\kappa)+\kappa\left(67\kappa^{4}+120\kappa^{2}+45\right)\right)}{5\left(3\left(3\kappa^{2}+10\right)\left(\kappa^{2}+1\right)^{2}\cot^{-1}(\kappa)+\kappa\left(31\kappa^{4}+59\kappa^{2}+30\right)\right)^{2}},\\
\tilde{I}_{2} & = & \frac{32\kappa^{4}\left(15\left(6\kappa^{2}+13\right)\left(\kappa^{2}+1\right)^{2}\cot^{-1}(\kappa)+\kappa\left(222\kappa^{4}+415\kappa^{2}+195\right)\right)}{5\left(3\left(3\kappa^{2}+10\right)\left(\kappa^{2}+1\right)^{2}\cot^{-1}(\kappa)+\kappa\left(31\kappa^{4}+59\kappa^{2}+30\right)\right)^{2}},\\
\tilde{I}_{3} & = & \frac{64\kappa^{6}\left(15\left(9\kappa^{2}+23\right)\left(\kappa^{2}+1\right)^{2}\cot^{-1}(\kappa)+\kappa\left(377\kappa^{4}+710\kappa^{2}+345\right)\right)}{5\left(\kappa^{2}+1\right)\left(3\left(3\kappa^{2}+10\right)\left(\kappa^{2}+1\right)^{2}\cot^{-1}(\kappa)+\kappa\left(31\kappa^{4}+59\kappa^{2}+30\right)\right)^{2}}.\end{eqnarray}
The differential equations now read\begin{eqnarray}
\partial_{\kappa}\left[\kappa^{2}\Lambda\right] & = & \left(\kappa^{2}\Lambda\right)^{2}\tilde{I}_{1}+\kappa^{2}\Lambda\left(\frac{2}{\kappa}+\tilde{I}_{2}\right)+\frac{1}{4}\tilde{I}_{3}\label{eq:Lamda_as}\\
\partial_{\kappa}\left[\kappa^{2}\Lambda^{\prime}\right] & = & \left(\kappa^{2}\Lambda^{\prime}\right)^{2}\tilde{I}_{1}+\kappa^{2}\Lambda^{\prime}\left(\frac{2}{\kappa}-2\tilde{I}_{2}\right)+\tilde{I}_{3}\label{eq:Lamdap_as}\end{eqnarray}
The $\tilde{I}$'s have simple asymptotic expansions for large $\kappa$,\begin{eqnarray}
\tilde{I}_{1} & = & \frac{28}{125}\kappa^{-1},\\
\tilde{I}_{2} & = & \frac{156}{125}\kappa^{-1},\\
\tilde{I}_{3} & = & \frac{512}{125}\kappa^{-1}.\end{eqnarray}
The asymptotic solution to the equation for $\Lambda$ (\ref{eq:Lamda_as})
can be obtained using these results as\begin{equation}
\Lambda(\kappa)\sim-\frac{1}{4}\left[29-5\sqrt{\frac{215}{7}}\frac{1-c\,\kappa^{\frac{2}{5}\sqrt{\frac{301}{5}}}}{1+c\,\kappa^{\frac{2}{5}\sqrt{\frac{301}{5}}}}\right].\end{equation}
\begin{figure}
\begin{centering}
\includegraphics[width=8cm]{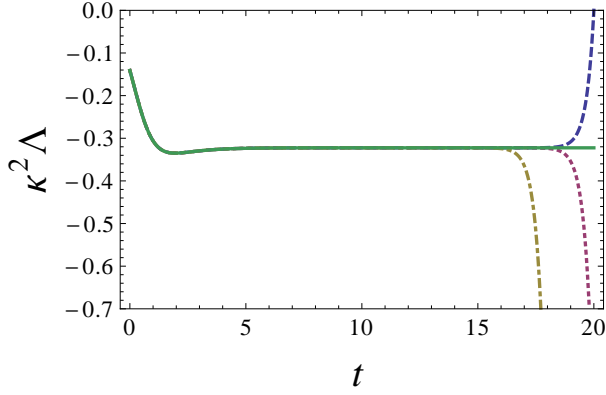}
\end{centering}
\caption{{(}Colour online{)} Four examples of the evolution of $\kappa^{2}\Lambda$. These were
solved starting the integration at $t=\ln\kappa=20$. The three
dashed lines represent the solutions for three random choices of the
initial value. The solid (green) line starts using Eq.~(\ref{eq:lam_as_c0})
as initial condition. \label{fig:lambda} }
\end{figure}
It is very tempting to assume $c\neq0$, and get the asymptotic solution
for $\kappa\rightarrow\infty$\begin{eqnarray}
\kappa^{2}\Lambda(\kappa) & \sim & -\frac{1}{4}\left(29+5\sqrt{\frac{215}{7}}\right).\label{eq:lam_as_large}\end{eqnarray}
Numerical investigations, Fig.~\ref{fig:lambda}, of the solution
to the {}``full'' equation, where we have not made the asymptotic
expansion, starting the integration from $t=\ln(\kappa)=20$, shows
that when we have chosen an arbitrary initial condition at finite
$\kappa$, the solution typically has a quick decay to the fixed point
value for $c=0$, for almost all initial conditions. Only when we
start very close to the asymptotic value (\ref{eq:lam_as_large})
do we find an apparently divergent numerical solution. Thus it is
actually most efficient to use the asymptotic solution with the choice
$c=0$, \begin{equation}
\kappa^{2}\Lambda(\kappa)\sim-\frac{1}{4}\left(29-5\sqrt{\frac{215}{7}}\right),\label{eq:lam_as_c0}\end{equation}
as initial condition for large $\kappa$.

The differential equation for $\Lambda^{\prime}$ does not have such
a trivial fixed point; the asymptotic solution is found to be quasi-periodic
\begin{equation}
\kappa^{2}\Lambda^{\prime}(\kappa)=\frac{1}{28}\left(31+5\sqrt{535}\tan\left(\frac{1}{25}\sqrt{535}\ln(\kappa)+\phi_{0}\right)\right).\label{eq:kappa2Lambdap}\end{equation}
This limit cycle-behaviour is the basic manifestation of the Efimov
effect in the approach employed here.

\subsubsection{Four-body couplings}

For the fur body-couplings we can also define scaled integrals,
\begin{eqnarray}
\tilde{K}_{1} & = & \frac{192\pi^{2}\kappa^{4}\left(\kappa^{2}+1\right)^{3}}{5\left(3\left(\kappa^{2}+1\right)^{2}\cot^{-1}(\kappa)+\kappa\left(5\kappa^{2}+3\right)\right)}\nonumber \\
 &  & \times\frac{\left(15\left(\kappa^{2}+1\right)^{3}\cot^{-1}(\kappa)+\kappa\left(17\kappa^{4}+40\kappa^{2}+15\right)\right)}{\left(3\left(\kappa^{2}+1\right)^{2}\left(\kappa^{2}+8\right)\cot^{-1}(\kappa)+\kappa\left(21\kappa^{4}+43\kappa^{2}+24\right)\right)^{2}},\\
\tilde{K}_{2} & = & \frac{4\kappa^{4}}{3\pi^{2}\left(\kappa^{2}+1\right)^{3}},\\
\tilde{K}_{3} & = & \frac{8\kappa^{4}}{3\pi^{2}\left(\kappa^{2}+1\right)^{4}}.
\end{eqnarray}
We can rewrite the differential equations as
\begin{align}
\partial_{\kappa}\bar{U}_{2} & =\frac{1}{2}\bar{U}_{2}^{2}\tilde{K}_{1}-2\Lambda^{\prime}\tilde{K}_{2}+\frac{3}{2}\tilde{K}_{3},\\
\partial_{\kappa}U_{2} & =\frac{1}{2}U_{2}^{2}\tilde{K}_{1}-2\Lambda\tilde{K}_{2}-\frac{3}{4}\tilde{K}_{3},\end{align}
and the asymptotic expansions\begin{eqnarray}
\tilde{K}_{1} & = & \frac{4\pi^{2}\kappa^{2}}{15},\\
\tilde{K}_{2} & = & \frac{4}{3\pi^{2}\kappa^{2}},\\
\tilde{K}_{3} & = & \frac{8}{3\pi^{2}\kappa^{4}}.\end{eqnarray}
It pays off to use \begin{equation}
V_{\boson\boson}(\kappa)=\kappa^{3}\bar{U}_{2}\end{equation}
This satisfies\begin{equation}
\kappa\partial_{\kappa}V_{\boson\boson}=\frac{2\pi^{4}}{15}V_{\boson\boson}(\kappa)^{2}+3V_{\boson\boson}(\kappa)+\frac{4}{\pi^{2}}-\frac{8}{5\pi^{2}}\kappa^{2}\Lambda^{\prime}(\kappa).\end{equation}
Without the additional $\Lambda$ term we find for large $\kappa$
that\begin{equation}
V_{\boson\boson}\rightarrow-\frac{45+\sqrt{1545}}{4\pi^{2}}\approx-2.13551.\end{equation}
By numerical solution we find that the periodic oscillations in $\Lambda$
dominate over this behaviour.

\subsection{Scattering lengths}

\subsubsection{Dimer-dimer scattering}

Our effective action contains the term \begin{align}
  V_{\boson\boson}= & \frac{1}{2}u_{2}\int\delta^{(4)}(p_{1}+p_{3}-p_{2}-p_{4})\left[\phi_{\alpha}^{\dagger}(p_{10},\vec{p}_{1})\phi_{\alpha}(p_{20},\vec{p}_{2})\right]\left[\phi_{\beta}^{\dagger}(p_{30},\vec{p}_{3})\phi_{\beta}(p_{40},\vec{p}_{4})\right]\nonumber \\
  & +\frac{1}{12}u_{2}^{'}\int\delta^{(4)}(p_{1}+p_{2}-p_{3}-p_{4})\left[\vvc^{\dagger}(p_{10},\vec{p}_{1})\cdot\vvc^{\dagger}(p_{20},\vec{p}_{2})-\ivc^{\dagger}(p_{10},\vec{p}_{1})\cdot\ivc^{\dagger}(p_{20},\vec{p}_{2})\right]\nonumber \\
  &
  \qquad\times\left[\vvc(p_{30},\vec{p}_{3})\cdot\vvc(p_{40},\vec{p}_{4})-\ivc(p_{30},\vec{p}_{3})\cdot\ivc(p_{40},\vec{p}_{4})\right].\end{align}
This is in terms of classical fields; the corresponding operator in
the {}``pole approximation'' when the cut-off reaches zero
is
\begin{align}
  \hat{V}_{\boson\boson}= & \frac{1}{2}u_{2}^{\prime}:\hat{n}^{2}:\nonumber \\
  & +\frac{1}{12}u_{2}\left[\hat{\vvc}^{\dagger}(E_{B},0)\cdot\hat{\vvc}^{\dagger}(E_{B},0)-\hat{\ivc}^{\dagger}(E_{B},0)\cdot\hat{\ivc}^{\dagger}(E_{B},0)\right]\\
&\quad
\left[\hat{\vvc}(E_{B},0)\cdot\hat{\vvc}(E_{B},0)-\hat{\ivc}(E_{B},0)\cdot\hat{\ivc}(E_{B},0)\right]\nonumber \\
  = & \frac{1}{2}u_{2}^{\prime}\hat{n}\left(\hat{n}-1\right)\nonumber \\
  &
  +\frac{1}{12}u_{2}\left[\hat{\vvc}^{\dagger}(E_{B},0)\cdot\hat{\vvc}^{\dagger}(E_{B},0)-\hat{\ivc}^{\dagger}(E_{B},0)\cdot\hat{\ivc}^{\dagger}(E_{B},0)\right]\\
&\quad
\left[\hat{\vvc}(E_{B},0)\cdot\hat{\vvc}(E_{B},0)-\hat{\ivc}(E_{B},0)\cdot\hat{\ivc}(E_{B},0)\right].\end{align}
Note the important effect of normal ordering; this is simply a
re-statement of the fact that $u_{2}^{\prime}$ does not contribute in
the single dimer channels. We now calculate the scattering between the
normalised scalar ($[2]$ denotes the symmetric irrep of SU(4)
\cite{Wybourne}) states\begin{equation} \left|[2](0,0)\right\rangle
  =\frac{1}{\sqrt{6\times2}}\left[\hat{\vvc}^{\dagger}(E_{B},0)\cdot\hat{\vvc}^{\dagger}(E_{B},0)-\hat{\ivc}^{\dagger}(E_{B},0)\cdot\hat{\ivc}^{\dagger}(E_{B},0)\right]\left|0\right\rangle
  ,\end{equation} and we get\begin{equation} \left\langle
    [2](0,0)\right|V_{\boson\boson}\left|[2](0,0)\right\rangle
  =\bar{u}_{2}.\end{equation} We thus find that\begin{align}
  a_{\boson\boson} & =\frac{M_{B}}{4\pi}\bar{u}_{2}/Z_{\phi}^{2}=\frac{M_{{\fermion}}}{2\pi}\bar{u}_{2}/Z_{\phi}^{2}\nonumber \\
  & =a_{\vb}32\pi(\bar{U}_{2})\end{align} We also use the fact that
$Z_{\phi}(0)=a_{\vb}g^{2}M^{2}/8\pi$.

In the same vein we  find in the other channel ($[1,1]$ denotes
the antisymmetric irrep of SU(4)) \begin{equation}
\left|[1,1](0,0)\right\rangle =\frac{1}{\sqrt{6\times2}}\left[\hat{\vvc}^{\dagger}(E_{B},0)\cdot\hat{\vvc}^{\dagger}(E_{B},0)+\hat{\ivc}^{\dagger}(E_{B},0)\cdot\hat{\ivc}^{\dagger}(E_{B},0)\right]\left|0\right\rangle ,\end{equation}
that\begin{align}
a_{\boson\boson}^{AS} & =\frac{M_{B}}{4\pi}u_{2}/Z_{\phi}^{2}=\frac{M_{{\fermion}}}{2\pi}u_{2}/Z_{\phi}^{2}\nonumber \\
 & =a_{\vb}32\pi(U_{2})\end{align}
In the mean-field limit, we find that $\lim_{\kappa\rightarrow0}U_{2}=1/16\pi$,
and we get the standard result $a_{\boson\boson}^{AS}=2a_{\vb}$.

\subsubsection{Dimer-nucleon scattering}
There are two possible dimer-nucleon state we can construct, a
symmetric and antisymmetric one.

We first assume that the scattering is evaluated in the channel of the
symmetric dimer-nucleon states\begin{equation}
  \left|\boson\fermion\right\rangle
  =\frac{1}{\sqrt{2}}\left[\hat{\vb}^{\dagger}(E_{B},0)\hat{\psi}^{\dagger}(E_{B}/2,0)-\hat{\ib}^{\dagger}(E_{B},0)\hat{\psi}^{\dagger}(E_{B}/2,0)\right]_{m_{S},m_{\isum}}^{1/2,1/2}\left|0\right\rangle
  ,\end{equation} we get\begin{equation} \left\langle
    DF\right|\hat{\vb}\left|DF\right\rangle =\lambda.\end{equation} We
thus conclude
\begin{align}
a_{\boson\fermion} & =\frac{M_{red}}{4\pi}\lambda/Z_{\phi}=\frac{2M_{{\fermion}}/3}{4\pi}\lambda/Z_{\phi}\nonumber \\
 & =a_{\vb}\frac{4}{3}\Lambda.\end{align}
In the antisymmetric case, defined by\begin{equation}
\left|\boson\fermion^{\prime}\right\rangle =\frac{1}{\sqrt{2}}\left[\hat{\vb}^{\dagger}(E_{B},0)\hat{\psi}^{\dagger}(E_{B}/2,0)+\hat{\ib}^{\dagger}(E_{B},0)\hat{\psi}^{\dagger}(E_{B}/2,0)\right]_{m_{S},m_{\isum}}^{1/2,1/2}\left|0\right\rangle ,\end{equation}
we get \begin{align}
a_{\boson\fermion}^{\prime} & =\frac{M_{red}}{4\pi}\lambda^{\prime}/Z_{\phi}=\frac{2M_{{\fermion}}/3}{4\pi}\lambda^{\prime}/Z_{\phi}\nonumber \\
 & =a_{\vb}\frac{4}{3}\Lambda^{\prime}.\end{align}

\section{Link between $T$ and $V$}

Our main results are for the threshold behaviour of various quantities are, 
in the main, expressed in terms of scattering lengths. 
The relation between our effective action and a scattering length is
based on the definition
\begin{equation}
\lim_{k\rightarrow0}T(k)=\frac{2\pi}{M_{red}}a_{\vb}.\end{equation}
For identical mass, $M_{1}=M_{2}=M$, we get\begin{equation}
T(0)=\frac{4\pi}{M}a_{\vb}.\end{equation}
What we call a {}``potential'' in the effective action, is really
nothing but the classical form of the $T$ matrix. For dimers we find
a similar relation, but we have to interpret $T$ as the expectation
value of the quantised potential in the relevant channel. As shown
in the appendix, this allows us to take into account symmetry aspects
of the problem. Thus \begin{equation}
V=\left\langle \hat{V}\right\rangle =T_{\boson\boson}(0)=\frac{4\pi a_{{\boson}\boson}}{M_{{\boson}}},\end{equation}
which can be used to show that the dimer-dimer scattering length in
the mean-field limit goes to $2a_{\vb}$.

\end{document}